%% file: ms.tex
\documentclass[traditabstract]{aa} 
\usepackage{graphicx}
\usepackage{txfonts}
\newcommand{\bff}{}
\def\velo{{\rm v}}
\def\gtsima{$\;\buildrel > \over \sim \;$}
\def\simgt{\lower.5ex \hbox{\gtsima}}
\def\ltsima{$\;\buildrel < \over \sim \;$}
\def\simlt{\lower.5ex \hbox{\ltsima}}

\begin{document}

   \title{Sulphur-bearing molecules in diffuse molecular clouds: new results from SOFIA/GREAT and the IRAM 30 m telescope}
   \author{D.~A.~Neufeld\inst{1}, B.~Godard\inst{2,3}, M.~Gerin \inst{2,3}, G.~Pineau des For\^ets\inst{2,3,4}, C.~Bernier \inst{5}, E.~Falgarone \inst{2,3}, U.~U.~Graf\inst{6}, R.~G\"usten\inst{7}, E.~Herbst \inst{5}, P.~Lesaffre \inst{2,3},   P.~Schilke\inst{6},   P.~Sonnentrucker\inst{8},   and H.~Wiesemeyer\inst{6}}

   \institute{The Johns Hopkins University, 3400 North Charles St.\, Baltimore, MD 21218, USA \and
LERMA, Observatoire de Paris, PSL University, CNRS, UMR 8112, F-75014, Paris, France
\and Sorbonne Universit\'es, UPMC Univ. Paris 6, UMR 8112, LERMA, F-75005, Paris, France
\and Institut d'Astrophysique Spatiale, CNRS UMR 8617, Universit\'e Paris-Sud, Orsay, France 
\and Department of Chemistry, University of Virginia, McCormick Road, P.O. Box 400319,
Charlottesville, VA 22904, USA 
\and I. Physikalisches Institut der Universit\"at zu K\"oln, Z\"ulpicher Strasse 77, 50937 K\"oln, Germany
\and
 Max-Planck-Institut f\"ur Radioastronomie, Auf dem H\"ugel 69, 53121 Bonn, Germany
 \and Space Telescope Science Institute, 3700 San Martin Drive, Baltimore, MD 21218  }

  \abstract
{
We have observed five sulphur-bearing molecules in foreground diffuse molecular clouds lying along the sight-lines to five bright continuum sources.  We have used the GREAT instrument on SOFIA to observe the SH 1383~GHz $^2\Pi_{3/2} \, J = 5/2 \leftarrow 3/2$ lambda doublet towards the star-forming regions W31C, G29.96--0.02, G34.3+0.1, W49N and W51, detecting foreground absorption towards all five sources; and the EMIR receivers on the IRAM 30m telescope at Pico Veleta to detect the H$_2$S $1_{10} - 1_{01}$ (169~GHz), CS $J=2-1$ (98~GHz) and SO $3_2 - 2_1$ (99~GHz) transitions.  Upper limits on the $\rm H_3S^+$ $1_0-0_0$ (293~GHz) transitions were also obtained at the IRAM 30 m.  In nine foreground absorption components detected towards these sources, the inferred column densities of the four detected molecules showed relatively constant ratios, with $N({\rm SH})/N({\rm H_2S})$ in the range 1.1 -- 3.0, $N({\rm CS})/N({\rm H_2S})$ in the range 0.32 -- 0.61, and $N({\rm SO})/N({\rm H_2S})$ in the range 0.08 -- 0.30.  The column densities of the sulphur-bearing molecules are very well correlated amongst themselves, moderately well correlated with CH (a surrogate tracer for H$_2$), and poorly correlated with atomic hydrogen.  The observed SH/H$_2$ ratios -- in the range 5 to 26 $\times 10^{-9}$ -- indicate that SH (and other sulphur-bearing molecules) account for $\ll 1 \%$ of the gas-phase sulphur nuclei.  The observed abundances of sulphur-bearing molecules, however, greatly exceed those predicted by standard models of cold diffuse molecular clouds, providing further evidence for the enhancement of endothermic reaction rates by elevated temperatures or ion-neutral drift.  We have considered the observed abundance ratios in the context of shock and turbulent dissipation region (TDR) models.  {Using the TDR model, we find that the turbulent energy available at large scale in the diffuse ISM is sufficient to explain the observed column densities of SH and CS.   Standard shock and TDR models, however, fail to reproduce the column densities of H$_2$S and SO by a factor of about 10; more elaborate shock models -- in which account is taken of the velocity drift, relative to H$_2$, of SH molecules produced by the dissociative recombination of H$_3$S$^+$ -- reduce this discrepancy to a factor $\sim 3$.}}

   \titlerunning{Sulphur-bearing molecules}
	\authorrunning{Neufeld et al.}
   \maketitle
%

\section{Introduction}

\begin{table*}
\caption{List of observed sources}
\begin{tabular}{lcccccccccc}
\hline
\\
Source & R.A. & Dec & Distance$^a$ & 
${\rm v}_{\rm sys}$ & Date & 
\multicolumn{5}{c}{Integration time (s)} \\
	   & J2000 & J2000 & kpc & km/s & (SOFIA)          & SH    & H$_2$S & CS & SO & H$_3$S$^+$ \\
\hline
\\
W31C 				& 18:10:28.70 		& --19:55:50.0 & 4.95  &  --4 &  17 Jul 2013$^b$ &   677    & 8886 & 1831 & 3357 & 3697 \\
G29.96--0.02 	   	& 18:46:03.80       & --02:39:20.0 & 5.26  &  98 &  17 Jul 2013$^b$ &   752    & 9213 & 3682 & 3682 & 5531 \\
G34.3+0.1	   		& 18:53:18.70       &  +01:14:58.0 & 3.8   &  59 &  22 Jul 2013$^b$ &   677    & 9206 & 5495 & 5495 & 3711 \\
W49N 				& 19:10:13.20 		&  +09:06:12.0 & 11.11 &  8 &  28 Sep 2011$^c$ &   495    & 6400 & 1221 & 1221 & 2737\\
W51  				& 19:23:43.90  		&  +14:30:30.5 & 5.1   &  55 &  02 Nov 2013$^c$ &   752    & 6410 & 3371 & 3371 & 3039\\
\\
\hline

\end{tabular} 
\vskip 0.05 true in
\tablefoottext{a}{References for distance estimates in order listed:
Sanna et al. (2014); Zhang et al. (2014); Fish et al. (2003); Zhang et al. (2013); Sato et al. (2010).  All are trigonometric estimates except that for G34.3+0.1, which is a kinematic estimate.}

\tablefoottext{b}{SOFIA deployment from Christchurch, New Zealand}

\tablefoottext{c}{SOFIA deployment from Palmdale, CA, USA}

\end{table*}

\begin{table*}
\caption{List of observed spectral lines}
\begin{tabular}{llccccc}
\hline
\\
Molecule & Transition & Rest frequency & Beam size & $E_{\rm U}/k $  & $E_{\rm L}/k $ &  $A_{\rm UL}$ \\
	   &  & GHz & arcsec & K & K & s$^{-1}$ \\
\hline
\\
SH & $^2\Pi_{3/2} \, J = 5/2 \leftarrow 3/2$ $\,\,\,\,F=2^+ \leftarrow 2^-$ & 1382.9056 
& 21 & 66.4 & 0.0  & $0.47 \times 10^{-3}$  \\ 
 ...  & $^2\Pi_{3/2} \, J = 5/2 \leftarrow 3/2$ $\,\,\,\,F=3^+ \leftarrow 2^-$ & 1382.9101 
& ... & ... & ... & $4.72 \times 10^{-3}$  \\
 ...  & $^2\Pi_{3/2} \, J = 5/2 \leftarrow 3/2$ $\,\,\,\,F=2^+ \leftarrow 1^-$ & 1382.9168 
& ... & ... & ...  &$4.24 \times 10^{-3}$  \\
 ...  & $^2\Pi_{3/2} \, J = 5/2 \leftarrow 3/2$ $\,\,\,\,F=2^- \leftarrow 2^+$ & 1383.2365 
& ... & ... & ...  &$0.47 \times 10^{-3}$  \\
 ...  & $^2\Pi_{3/2} \, J = 5/2 \leftarrow 3/2$ $\,\,\,\,F=3^- \leftarrow 2^+$ & 1383.2412 
& ... & ... & ...  & $4.72 \times 10^{-3}$  \\
 ...  & $^2\Pi_{3/2} \, J = 5/2 \leftarrow 3/2$ $\,\,\,\,F=2^- \leftarrow 1^+$ & 1383.2478 
& ... & ... & ...  & $4.24 \times 10^{-3}$  \\
H$_2$S & $J_{K_a,K_c} = 1_{10} \leftarrow 1_{01}$ & \phantom{1}168.7628 & 14.6 & 27.88 & 19.88 & $2.68 \times 10^{-5}$  \\
CS &  $ J = 2 \leftarrow 1$ & \phantom{11}97.9810 & 25.1 & 7.05 & 2.35 & $1.67 \times 10^{-5}$ \\
SO &  $ J_N = 3_2 \leftarrow 2_1$ & \phantom{11}99.2999 & 25.6 & 9.23 & 4.46 & $1.13 \times 10^{-5}$ \\
$\rm H_3S^+$ & $K = 0$, $J = 1 \leftarrow0$ & \phantom{1}293.4572  & 8.4 & 14.08 & 0.00 & $2.94 \times 10^{-4}$ \\
\\
\hline
\end{tabular}
\end{table*}

\begin{table*}
\tiny
\centering
\caption{Continuum antenna temperatures and r.m.s.\ noise}
\begin{tabular}{llrcccc}

\hline
\\
Source & Line & & $T_A({\rm cont})$ & $T_{\rm rms}$ & Channel & S/N$^a$ \\
       &      & & (K) & (mK) & width  & ratio\\
       &      & &     &      & (km/s)\\
\\
\hline
\\
\input{rms.tex}
\hline
\tablefoottext{a}{Signal-to-noise ratio on the continuum}
\end{tabular}

\end{table*}
	   
To date, more than a dozen molecules containing sulphur have been detected unequivocally in interstellar gas clouds and/or circumstellar outflows: SH, SH$^+$, CS, NS, SO, SiS, H$_2$S, SO$_2$, HCS$^+$, OCS, C$_2$S, C$_3$S, HNCS, HSCN, H$_2$CS, and CH$_3$SH.  All but one of these molecules were first detected using ground-based telescopes operating at millimeter or submillimeter wavelengths, the only exception being the mercapto radical, SH, which has its lowest rotational transition at 1383~GHz, a frequency inaccessible from ground-based observatories and also lying in a gap between Bands 5 and 6 of the {\it Herschel Space Observatory}'s  
HIFI instrument.  The first detection of interstellar SH was obtained (Neufeld et al. 2012) in absorption towards the star-forming region W49N using the German Receiver for Astronomy at Terahertz Frequencies (GREAT; Heyminck et al.\ 2012)
instrument\footnote{GREAT is a development by the MPI f\"ür Radioastronomie and the KOSMA / Universit\"at zu K\"oln, in cooperation with the MPI f\"ur Sonnensystemforschung and the DLR Institut f\"ur Planetenforschung.} on the Stratospheric Observatory for Infrared Astronomy (SOFIA; Young et al.\ 2012).  

The chemistry of interstellar molecules containing sulphur is distinctive in that none of the species S, SH, S$^+$, SH$^+$, or H$_2$S$^+$ can react exothermically with H$_2$ in a hydrogen atom abstraction reaction (i.e.\ a reaction of the form $\rm X + H_2 \rightarrow XH + H$.)  This feature of the thermochemistry greatly inhibits the formation of sulphur-bearing molecules in gas at low temperature.

In this paper, we present a corrected analysis of the SH abundance presented in Neufeld et al. (2012), together with additional detections of SH obtained with SOFIA/GREAT toward four other bright continuum sources -- W51, W31C, G34.3+0.1 and G29.96--0.02 -- and ancillary observations of H$_2$S, CS, SO and H$_3$S$^+$ obtained with the IRAM 30~m telescope at Pico Veleta toward all five sources.  (The observations of H$_3$S$^+$ led only to upper limits, not detections.)  The observations and data reduction are described in \S 2 below, and results presented in \S 3.  The abundances of SH, H$_2$S, CS, and SO -- and upper limits on the H$_3$S$^+$ abundances --  derived in diffuse molecular clouds along the sight-lines to these bright continuum sources are discussed in \S 4 with reference to current astrochemical models. 

\section{Observations and data reduction}

Table 1 lists the continuum sources toward which SH, H$_2$S, CS, SO and H$_3$S$^+$ were observed, along with the exact positions targeted, the estimated distances to each source, the source systemic velocities relative to the Local Standard of Rest (LSR), the dates of each SOFIA observation, and the total integration time for each transition.  In Table 2, we list the transitions targeted for each species, together with their rest frequencies, telescope beam size at the transition frequency, upper and lower state energies, and spontaneous radiative decay rates.  All the molecular data presented here were taken directly from the CDMS (M\"uller et al.\ 2005) or JPL (Pickett et al.\ 1998) line catalogs, with the exception of those for the H$_3$S$^+$ transition; for the latter transition, which is not included in either catalog, we adopted the frequency measured by Tinti et al. (2006), and obtained the radiative decay rate for an assumed dipole moment of 1.731~D (Botschwina et al.\ 1986).  

All the SOFIA data reported here were obtained with the L1 receiver of GREAT and the AFFTS backend; the latter provides 8192 spectral channels with a spacing of 183.1~kHz.
As described in Neufeld et al. (2012), the W49N SH data were acquired in the upper sideband of the L1 receiver; observations of the other four sources were performed with the SH line frequency in the lower sideband.  At this frequency, the telescope beam has a diameter of $\sim 21^{\prime\prime}$ HPBW.  The observations were performed in dual beam switch mode, with a chopper frequency of 1 Hz and the reference positions located 60$^{\prime\prime}$ (75$^{\prime\prime}$ for W49N) on either side of the source along an east-west axis. 

The raw data were calibrated to the $T_A^*$ (``forward beam brightness temperature'') scale, using an independent fit to the dry and the wet content of the atmospheric emission.  Here, the assumed forward efficiency was 0.97 and the assumed main beam efficiency for the L1 band was 0.67 (except in the case of the W49N data from 2011, where the beam efficiency was 0.54).  The uncertainty in the flux calibration is estimated to be $\sim 20\%$ (Heyminck et al.\ 2012). 
Additional data reduction was performed using CLASS\footnote{Continuum and Line Analysis Single-dish Software.}.  This entailed the removal of a baseline and the rebinning of the data into 10-channel bins, resulting in a 0.40~km/s channel spacing for the final data product.  

All the IRAM data reported here were obtained using the EMIR receivers (Carter et al.\ 2012) on the 30-m telescope at Pico Veleta on 2014 March 16 -- 18, with the H$_2$S, CS, SO, and H$_3$S$^+$ transitions observed respectively using the E1, E0, E0, and E3 receivers.  Since two receivers can be used simultaneously (and upper and lower sideband spectra obtained separately and simultaneously for two orthogonal polarizations), we targeted the H$_2$S transition in all observations together with either (1) the H$_3$S$^+$ transition when the weather conditions were excellent (precipitable water vapor $< 1$~mm) or (2) the CS and SO transitions.  The FTS backend was deployed for all observations, providing a 200 kHz channel spacing over a 4.05 GHz bandwidth. In addition, the VESPA autocorrelator was used to target the CS transition toward W49N, W31C, G34.3+0.1 and G29.96--0.02 and the SO transition toward W49N and W31C; the VESPA backend was used to obtain a 80~kHz channel spacing over a bandwidth of 160 MHz.  All the data were acquired in double beam switching mode, using the wobbling secondary with a throw of $\pm 45$~arcsec.  Observations of Mars and Mercury were used for periodic checks of the telescope focus.

The CLASS software package was used to coadd all the data obtained for each source, with a weighting inversely proportional to the square of the radiometric noise in each scan.  The final coadded spectra show the antenna temperature, averaged over both polarizations, as a function of the frequency in the Local Standard of Rest (LSR) frame.

\begin{table*}
\caption{Derived molecular column densities}
\begin{tabular}{lcccccccc}
\hline
\\
Source & $\velo_{\rm LSR}$  & $N(\rm{SH})^a$ & $N(\rm{H_2S})$ & $N(\rm{CS})$ & $N(\rm{SO})$ &
$N(\rm{SH})/N(\rm H_2S)$ & $N(\rm{CS})/N(\rm H_2S)$ & $N(\rm{SO})/N(\rm H_2S)$ \\
& (km~s$^{-1}$) & ($10^{12}\,\rm{cm}^{-2}$) & ($10^{12}\,\rm{cm}^{-2}$) & ($10^{12}\,\rm{cm}^{-2}$) & ($10^{12}\,\rm{cm}^{-2}$)  \\
\\
\hline
\\
\input{column.tex}
\hline 
\end{tabular}
\tablefoottext{a}{Numbers in parentheses are 1 sigma statistical errors}

\tablefoottext{b}{Material absorbing in this velocity interval may lie in close proximity to W51, where radiative excitation is important: the derived column densities, computed for the conditions in foreground diffuse clouds, should therefore be regarded with caution.}
\end{table*}

\section{Results}
 
In Table 3, we report the continuum brightness temperatures measured at each frequency, together with the r.m.s noise in the coadded spectra.  The continuum spectra of H$_2$S, SO and CS are of a particularly high signal-to-noise ratio, providing exquisite sensitivity to weak spectral features.  In Figures 1 through 5, we show the spectra obtained toward each source.  The left panels are scaled to show the full range of observed antenna temperatures.  The H$_2$S, CS, and SO spectra show strong emission at the systemic velocity of each source, together with multiple narrower absorption features.  The SH spectra show only absorption.  The H$_3$S$^+$ spectra, which show multiple interloper emission lines (i.e. emission lines other than the targeted line), show no evidence for H$_3$S$^+$ emission or absorption.  The right panels show the same data with the vertical scaling adjusted to reveal weaker absorption features.

\begin{figure*}
\includegraphics[width=16 cm]{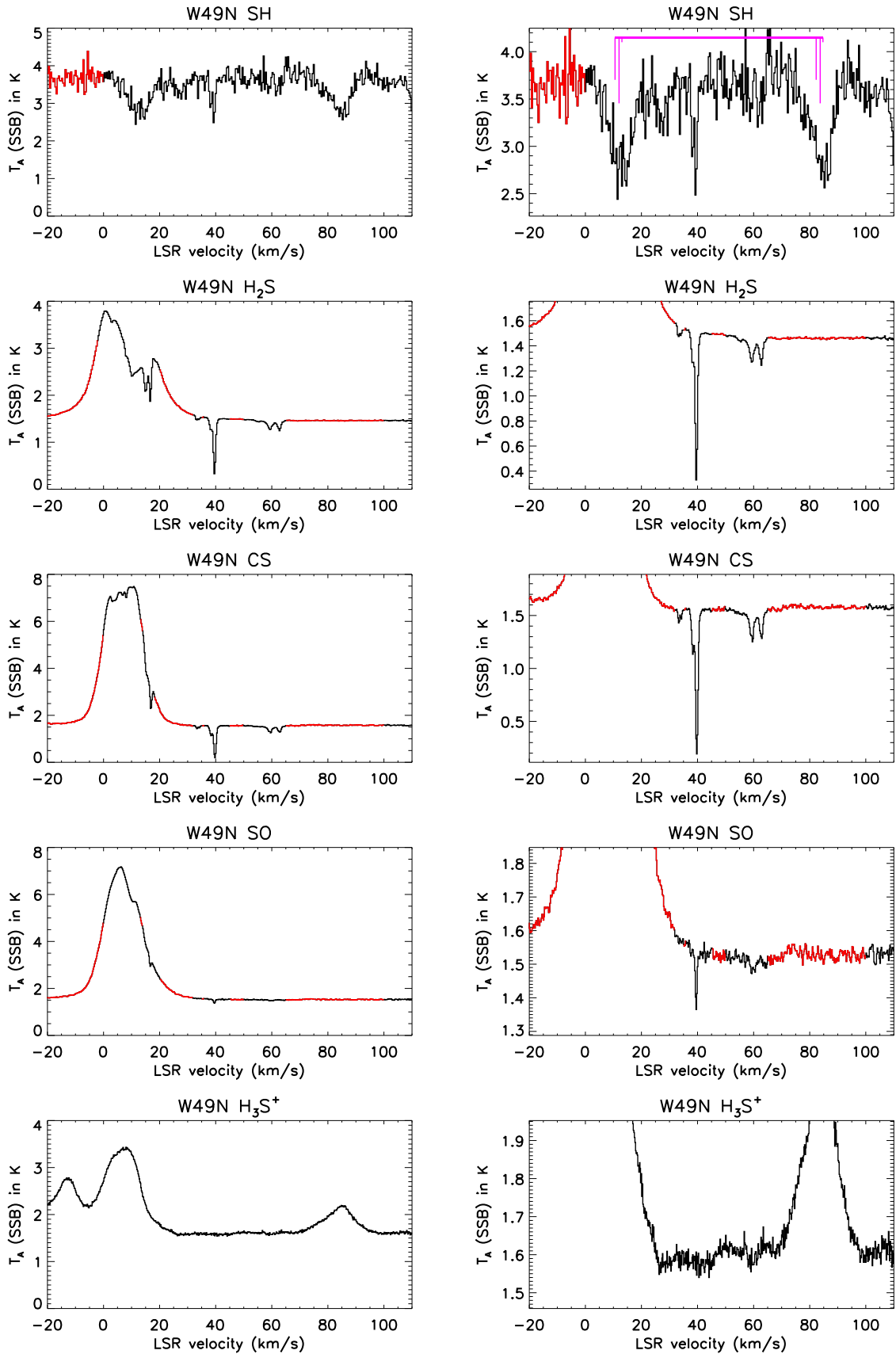}
\vskip 0.2 true in
\caption{Spectra obtained toward W49N.  Red sections of each plot indicate spectral regions assumed devoid of absorption for the purpose of fitting a continuum with emission lines.  The right panels are zoomed versions of the left panels, with the scaling chosen to reveal weak absorption features.  Magenta lines indicate the lambda doubling and hyperfine splittings for SH.}
\end{figure*}

\begin{figure*}
\includegraphics[width=16 cm]{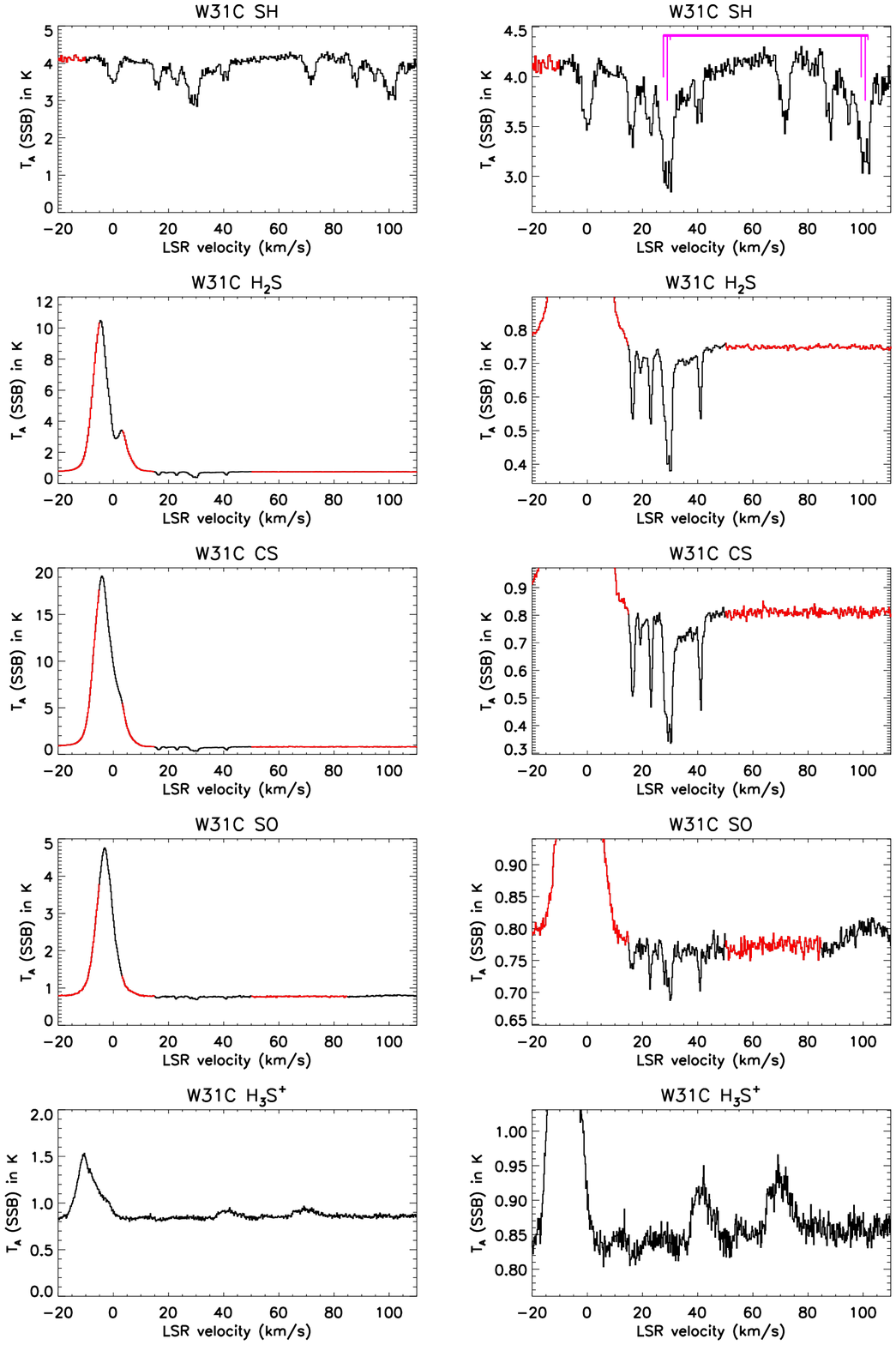}
\vskip 0.2 true in
\caption{Spectra obtained toward W31C.  Red sections of each plot indicate spectral regions assumed devoid of absorption for the purpose of fitting a continuum with emission lines. The right panels are zoomed versions of the left panels, with the scaling chosen to reveal weak absorption features.  Magenta lines indicate the lambda doubling and hyperfine splittings for SH.}
\end{figure*}

\begin{figure*}
\includegraphics[width=16 cm]{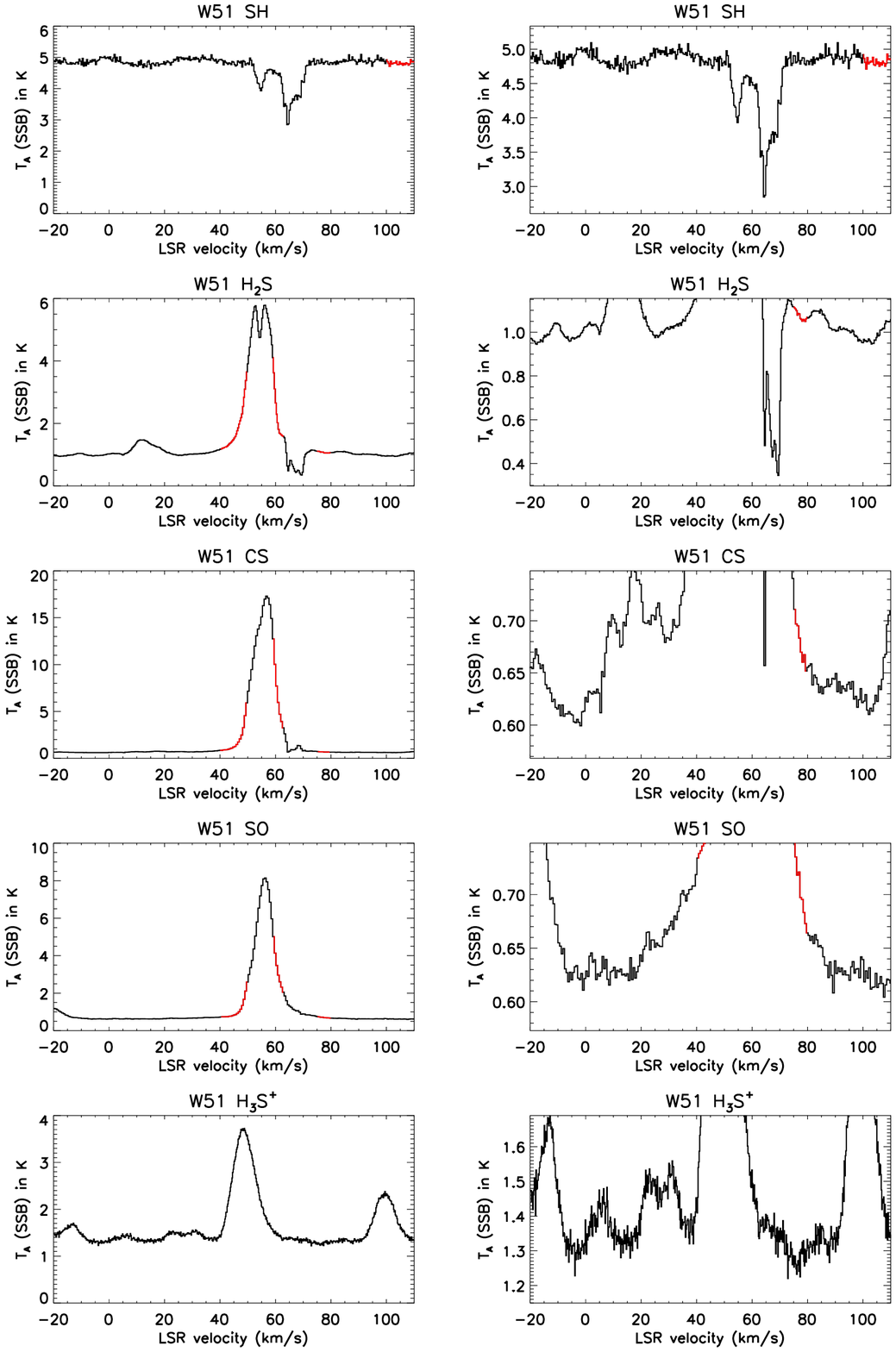}
\vskip 0.2 true in
\caption{Spectra obtained toward W51.  Red sections of each plot indicate spectral regions assumed devoid of absorption for the purpose of fitting a continuum with emission lines.  The right panels are zoomed versions of the left panels, with the scaling chosen to reveal weak absorption features.}
\end{figure*}

\begin{figure*}
\includegraphics[width=16 cm]{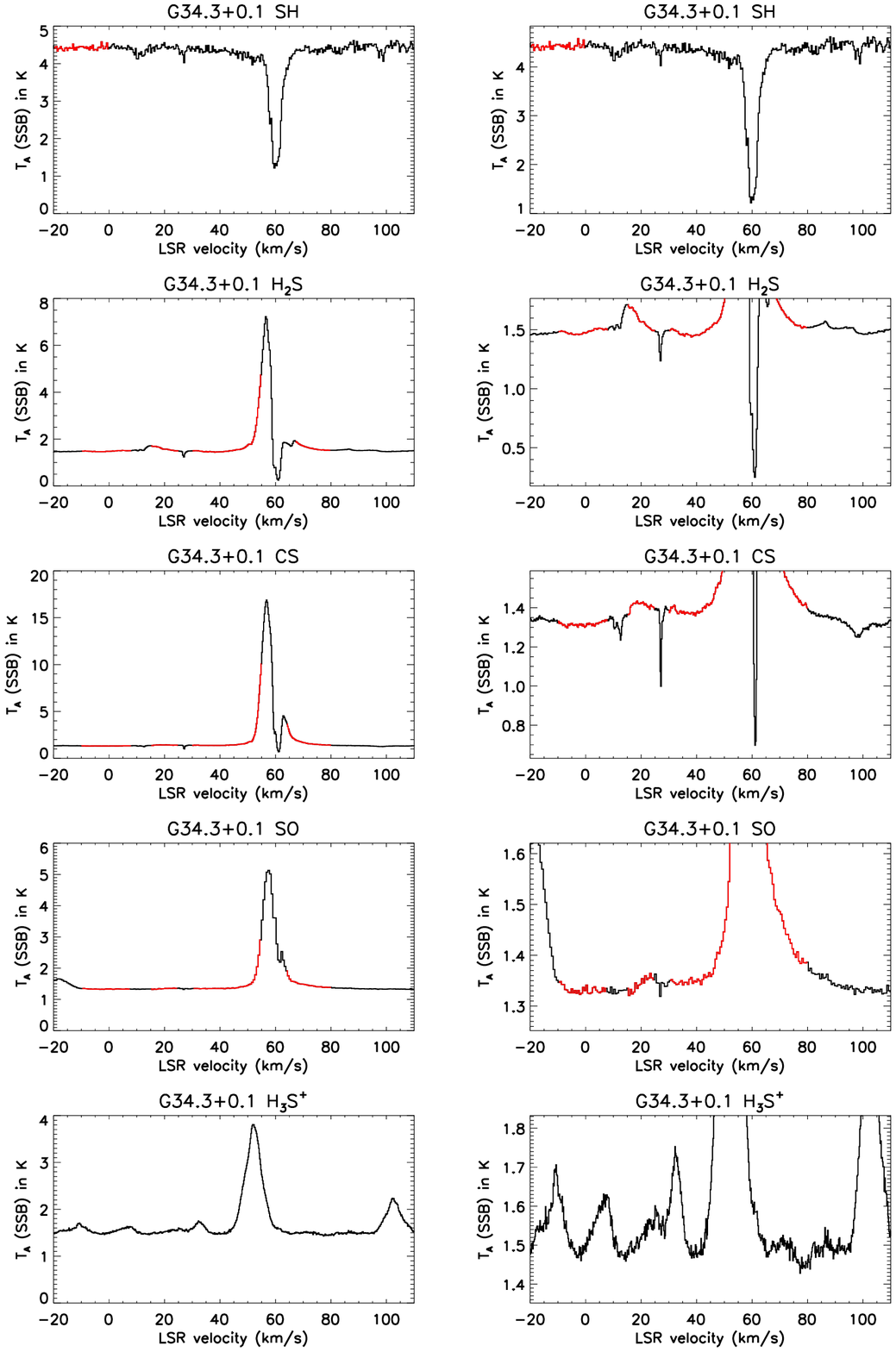}
\vskip 0.2 true in
\caption{Spectra obtained toward G34.3+0.1.  Red sections of each plot indicate spectral regions assumed devoid of absorption for the purpose of fitting a continuum with emission lines.  The right panels are zoomed versions of the left panels, with the scaling chosen to reveal weak absorption features.}
\end{figure*}

\begin{figure*}
\includegraphics[width=16 cm]{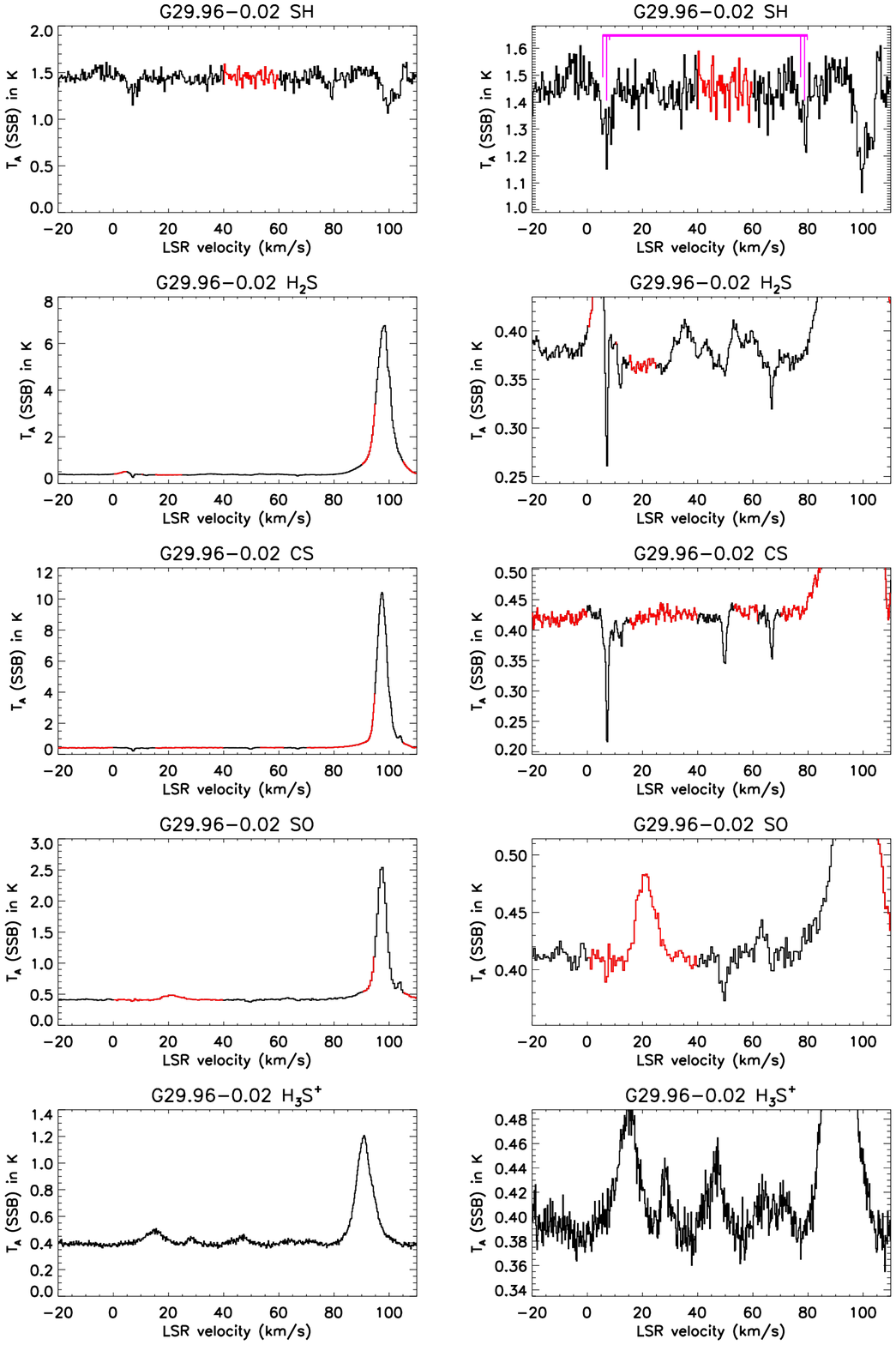}
\vskip 0.2 true in
\caption{Spectra obtained toward G29.96-0.02.  Red sections of each plot indicate spectral regions assumed devoid of absorption for the purpose of fitting a continuum with emission lines. The right panels are zoomed versions of the left panels, with the scaling chosen to reveal weak absorption features.  Magenta lines indicate the lambda doubling and hyperfine splittings for SH.}
\end{figure*}

\begin{table*}
\tiny
\caption{Derived molecular abundances}
\begin{tabular}{lccccccccccc}
\hline
\\
Source & $\velo_{\rm LSR}$  & $N(\rm{H_2})^a$ & $N({\rm H})^b$ &
$N(\rm{SH})/$ & $N(\rm H_2S)/$ &
$N(\rm{CS})/$ & $N(\rm{SO})/$ &
$N(\rm{SH})/$ & $N(\rm H_2S)/$ &
$N(\rm{CS})/$ & $N(\rm{SO})/$ \\
& (km~s$^{-1}$) & ($10^{21}\,\rm{cm}^{-2}$) & ($10^{21}\,\rm{cm}^{-2}$) & $N({\rm H}_2)^c$ & $N({\rm H}_2)$ & $N({\rm H}_2)$ & $N({\rm H}_2)$ & $N_{\rm H}$ & $N_{\rm H}$ & $N_{\rm H}$ & $N_{\rm H}$\\

&&&& $\times \, 10^{9}$ &  $\times \, 10^{9}$ &  $\times \, 10^{9}$ &  $\times \, 10^{9}$ &  $\times \, 10^{9}$ &  $\times \, 10^{9}$ &  $\times \, 10^{9}$ &  $\times \, 10^{9}$\\
\hline
\\
\input{abundance.tex}
\hline 
\end{tabular}
\vskip 0.1 true in
\tablefoottext{a}{Based on the CH column densities reported by Godard et al.\ (2012), for an assumed CH/H$_2$ ratio of $3.5 \times 10^{-8}$ (Sheffer et al.\ 2008), or - in the case of G29.96--0.02 - on the HF column density reported by Sonnentrucker et al.\ (2015), for an assumed HF/H$_2$ ratio of $1.4 \times 10^{-8}$}

\tablefoottext{b}{Obtained from 21 cm observations performed with the JVLA (Winkel et al.\ 2015) or the VLA (Fish et al.\ 2003, for the case of G29.96--0.02)} 

\tablefoottext{c}{The errors in the derived abundances are dominated by uncertainties in the CH/H$_2$ conversion factor (which is uncertain by a factor $\sim 1.6$)}
\end{table*} 

In each panel in Figures 1 -- 5, the red parts of each histogram denote spectral regions that were assumed to be devoid of foreground absorption.  These regions were subsequently fit with a flat continuum and one or more Gaussian emission features, to provide a background spectrum of the source itself, $T_{\rm bg} (\velo_{\rm LSR}) $.   The observed spectra were then divided by that fitted background spectrum (including emission lines), to obtain a transmission spectrum, $t = T_A / T_{\rm bg} $,  and an optical depth spectrum, $\tau (\velo_{\rm LSR})=-{\rm ln} \,t(\velo_{\rm LSR})$.  In the case of the SH transition, which is split by lambda doubling and hyperfine splitting, the velocity scale is computed for the $^2\Pi_{3/2} \, J = 5/2 \leftarrow 3/2$ $\,\,\,\,F=3^- \leftarrow 2^+$ transition.\footnote{With the rest frequencies listed in Table 2, the velocity centroids of the SH absorption features were found to be systematically offset from those of other molecules.  To bring them into accord, we adopted a +2 MHz velocity shift for all the SH frequencies in Table 2; a correction of this magnitude is consistent with the experimental uncertainties in the laboratory spectroscopy.  Figures 1 -- 5 include this frequency correction.}   As is clear from Figures 1 -- 5, the absorption pattern is therefore repeated with a velocity shift of +71.8~km/s, corresponding to the magnitude of the lambda doubling, and is broadened by the hyperfine structure. 

\begin{figure*}
\centering
\includegraphics[width=18 cm]{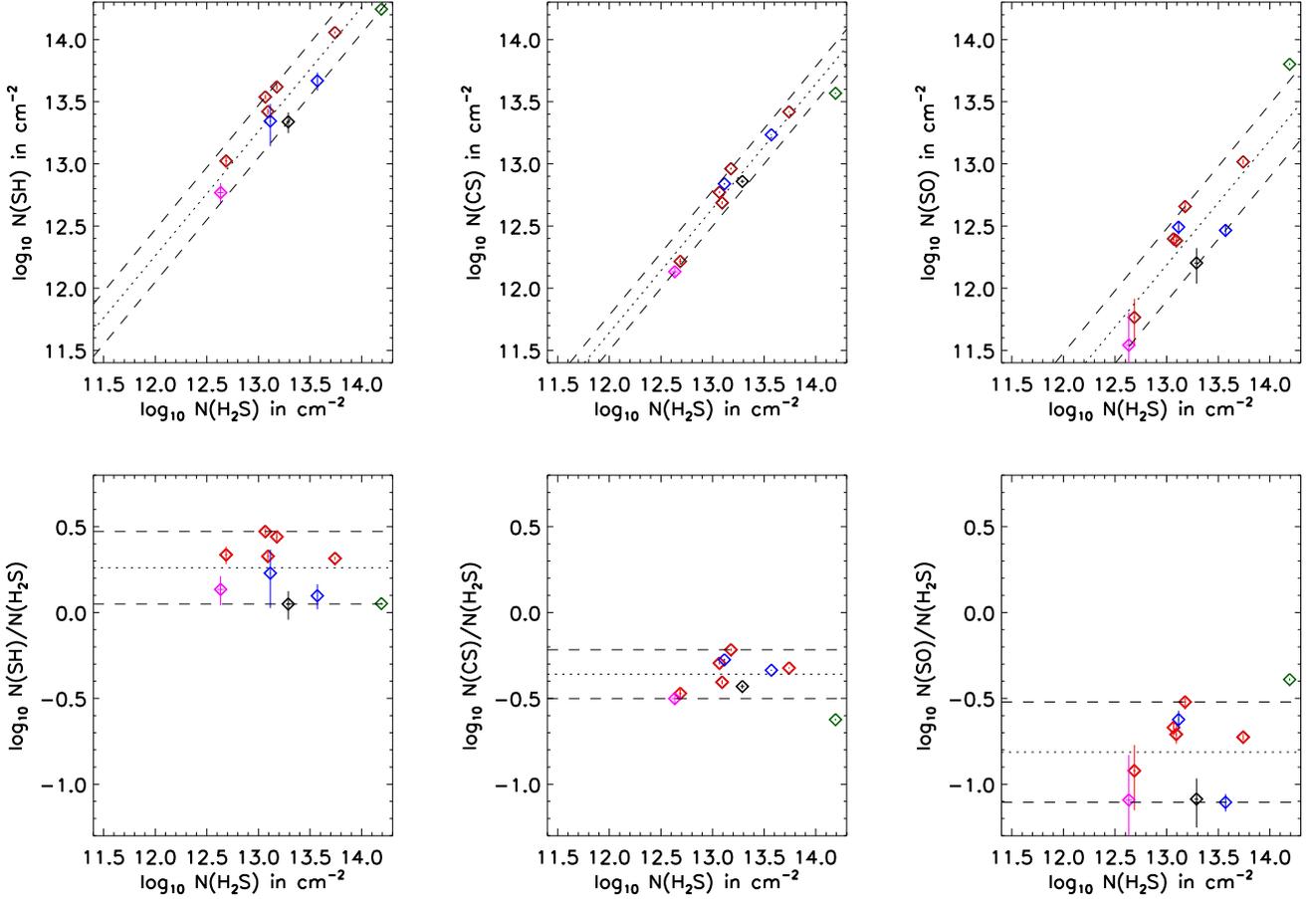}
\vskip 0.2 true in
\caption{Molecular column densities (top) and column density ratios (bottom) in foreground clouds along the sight-lines to W49N (blue), W31C (red), G34.3+0.1 (magenta) and G29.96--0.02 (black).  The error bars indicate $1\,\sigma$ statistical errors only. Dashed curves indicate the regions spanned by the range of column density ratios measured in this set of foreground clouds, with the dotted curve plotted midway between the dashed curves.  The green point applies to the 62 -- 75 km~s$^{-1}$ velocity range in W51, which probably lies in close proximity to the background continuum source. }
\end{figure*}

\begin{figure*}
\centering
\includegraphics[width= 13 cm]{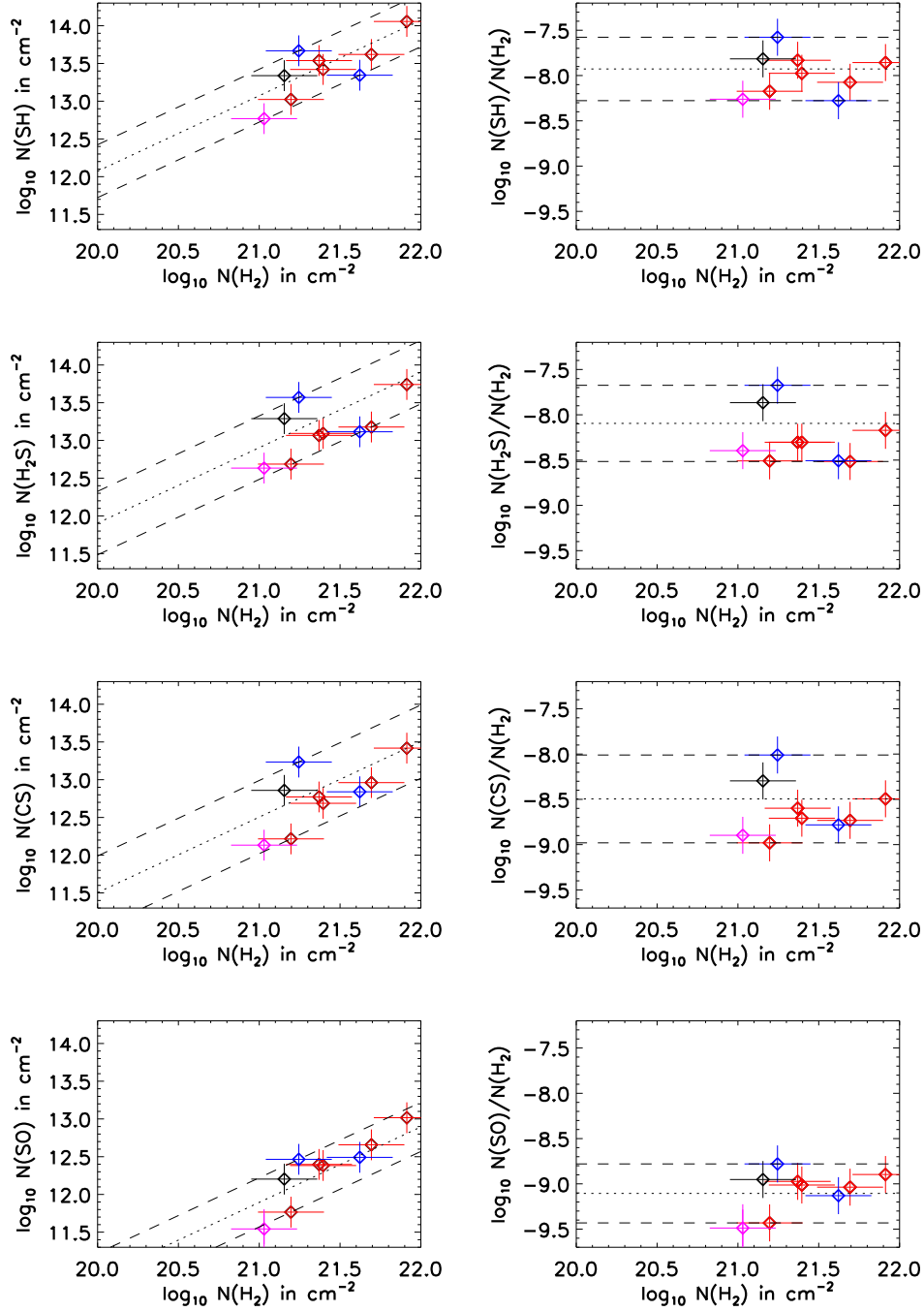}
\vskip 0.2 true in
\caption{Molecular column densities (left) and abundances relative to H$_2$ (right) in foreground clouds along the sight-lines to W49N (blue), W31C (red), G34.3+0.1 (magenta) and G29.96--0.02 (black).  The H$_2$ column densities are derived from CH column densities presented by Godard et al. (2012), for an assumed CH/H$_2$ ratio of $3.5 \times 10^{-8}$ (Sheffer et al. 2008).  The error bars are dominated by uncertainties in the CH to H$_2$ conversion, except for the SO values toward G34.3+0.1.
 Dashed curves indicate the regions spanned by the range of abundances measured in this set of foreground clouds, with the dotted curve plotted midway between the dashed curves.}
\end{figure*}

\begin{figure*}
\centering
\includegraphics[width=13 cm]{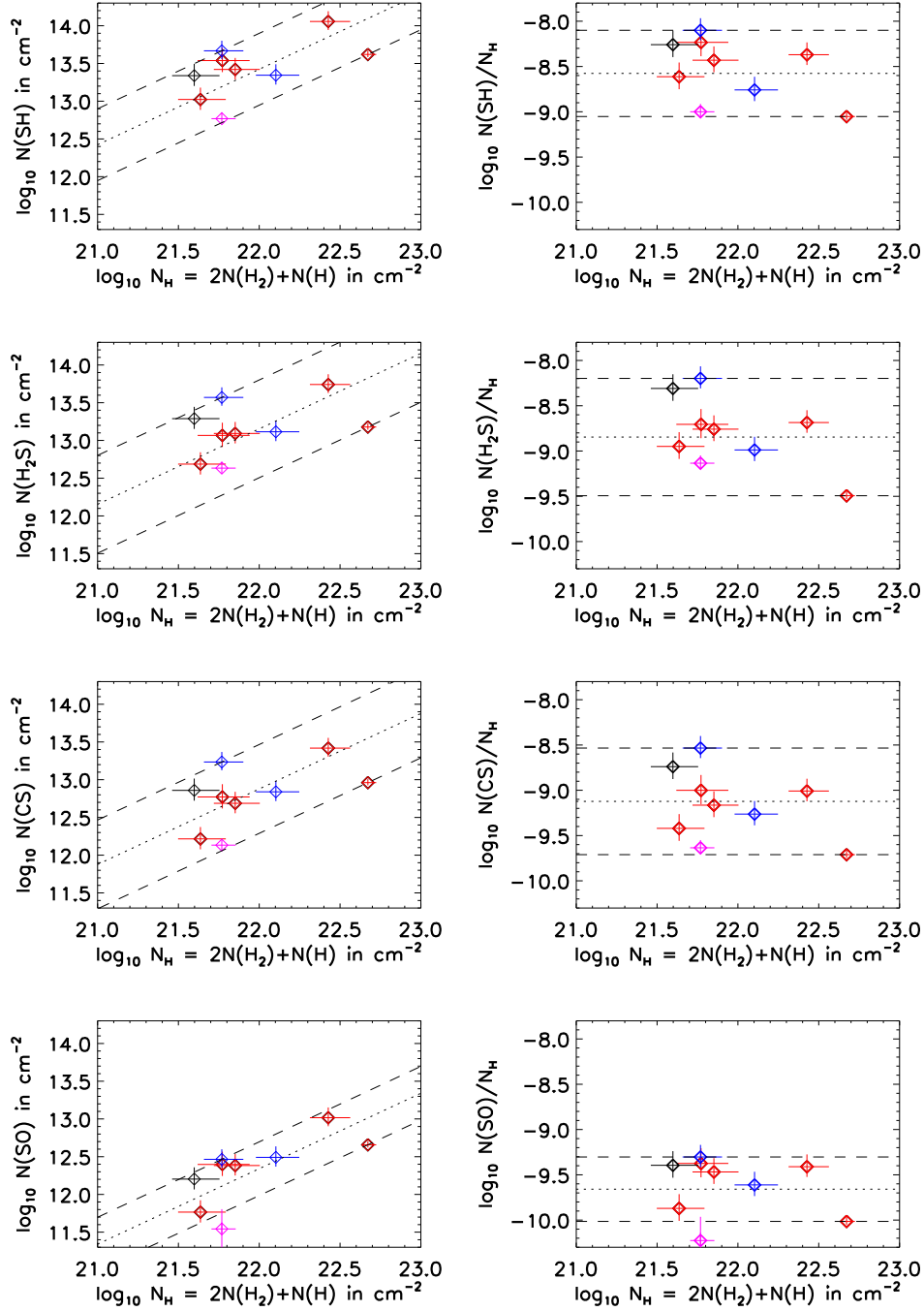}
\vskip 0.2 true in
\caption{Molecular column densities (left) and abundances relative to H nuclei (right) in foreground clouds along the sight-lines to W49N (blue), W31C (red), G34.3+0.1 (magenta) and G29.96--0.02 (black).  The H$_2$ column densities are derived from CH column densities presented by Godard et al. (2012), for an assumed CH/H$_2$ ratio of $3.5 \times 10^{-8}$ (Sheffer et al. 2008), and the atomic H column densities are derived from 21 cm observations reported by Winkel et al.\ (2015) or (for G29.96--0.02) by Fish et al.\ (2003).  The error bars are dominated by uncertainties in the CH to H$_2$ conversion, except for the SO values toward G34.3+0.1.  Dashed curves indicate the regions spanned by the range of abundances measured in this set of foreground clouds, with the dotted curve plotted midway between the dashed curves.}
\end{figure*}

\begin{figure}
\includegraphics[width=9 cm]{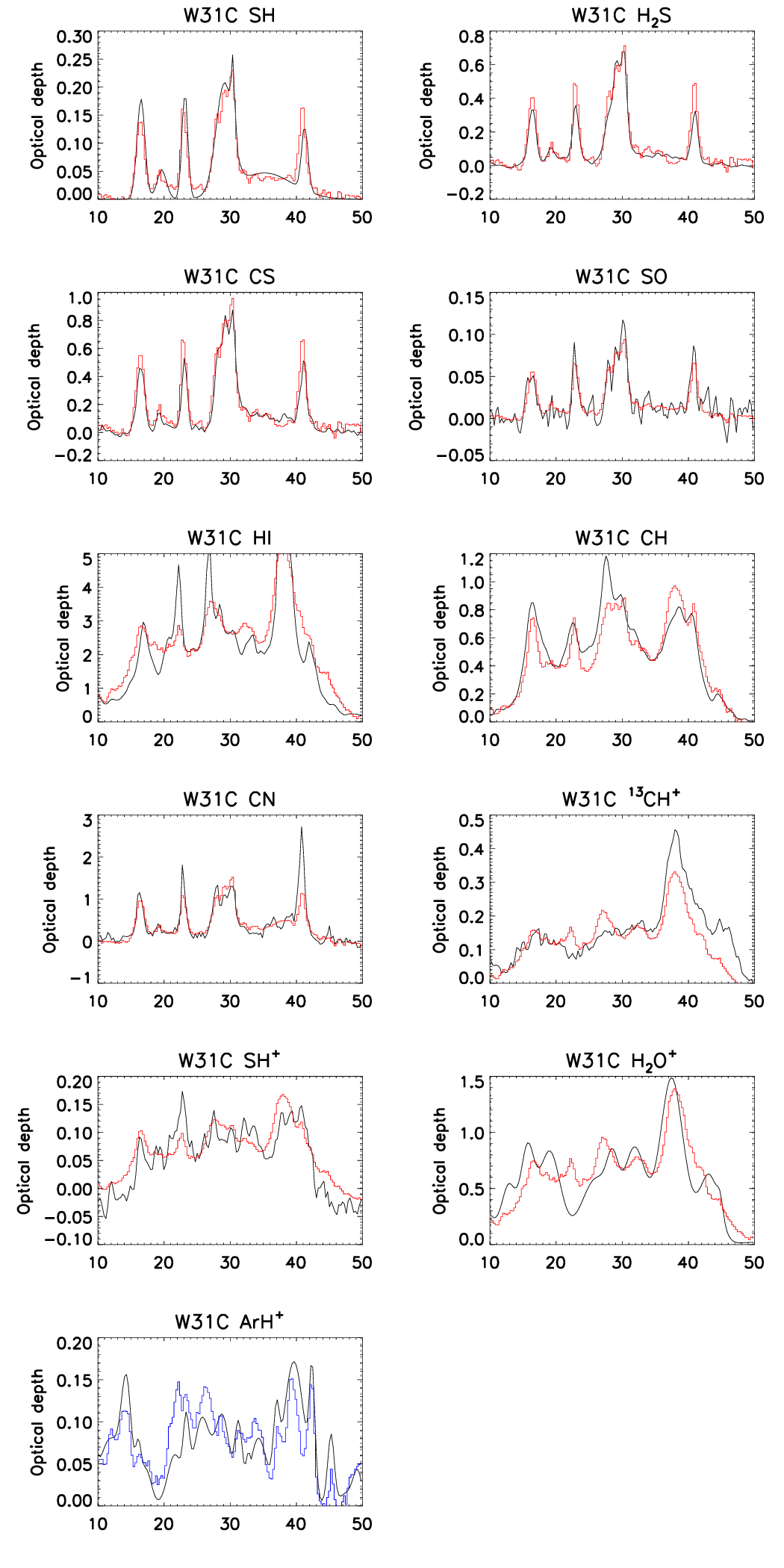}
\vskip 0.2 true in
\caption{Optical depths for eleven transitions observed toward W31C as a function of LSR velocity.  Black curve: rebinned observed data (see text).  Red histogram: approximate fit using only the first two principal components.  Blue histogram:  approximate fit using only the first three principal components.}
\end{figure}

\begin{figure}
\includegraphics[width=9 cm]{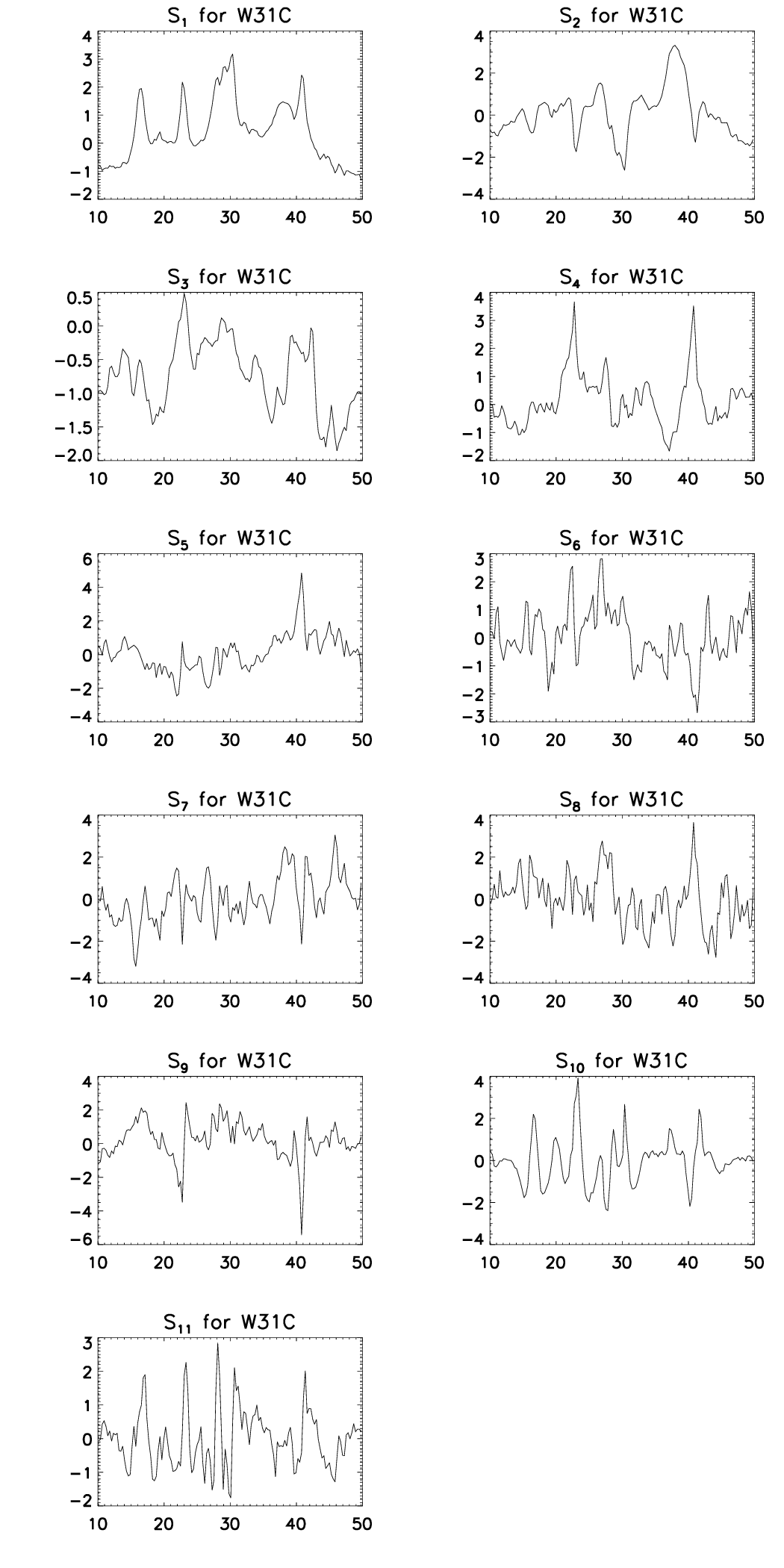}
\vskip 0.2 true in
\caption{Principal components obtained for W31C, from a joint analysis of the W31C and W49N absorption spectra}
\end{figure}

\begin{figure}
\includegraphics[width=9 cm]{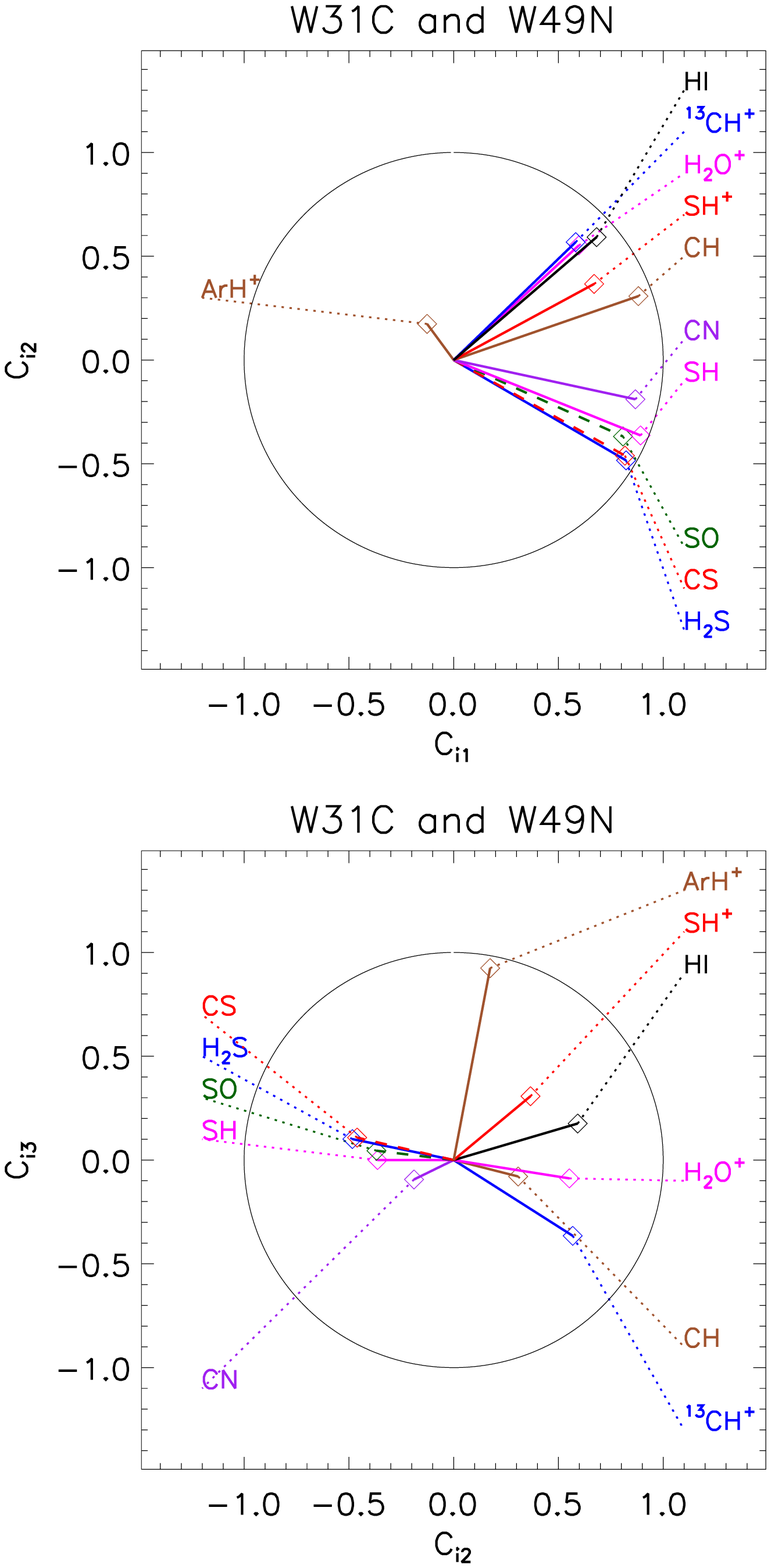}
\vskip 0.2 true in
\caption{Coefficients for the first two principal components obtained for each of the eleven transitions.  Solid lines connect the origin to the point representing the coefficients.  The black circles shows the unit circle within which all the coefficients must lie.}
\end{figure}

Figures A.1 through A.5 (black histogram), in Appendix A, show the optical depth as a function of LSR velocity for the SH, H$_2$S, CS, and SO transitions.  The vertical axis shows the total molecular column density per velocity interval.  The latter is computed for an excitation temperature of 2.73~K, the temperature of the cosmic microwave background. The validity of this assumption is justified by excitation calculations described in Appendix B.
In the case of H$_2$S, an ortho-to-para ratio of 3 was assumed.    Blue curves show multigaussian fits to the column density per velocity interval.  In the case of SH, the conversion to column density is complicated by the presence of hyperfine structure and lambda doubling.  Here, the blue curve represents the fitted column density per unit velocity interval, while the red curve represents the corresponding optical depth (after convolution with the hyperfine structure and lambda doubling.)  Thus, in this case, it is the {\it red} curve that should be compared with the optical depths shown in the black histogram.
The blue symbols above the spectra indicate the centroids and widths (full-widths-at-half-maximum) for each fitted Gaussian.  In our fits, the linewidths were constrained to be identical for all four molecules, and the centroids allowed to vary by at most 0.2~km/s.

In Table 4, we present the foreground molecular column densities inferred for selected velocity intervals toward each source, along with column density ratios.  The selected velocity windows are marked by the red vertical lines in Figures A.1 -- A.5, and the numbers appearing above each window give the molecular column densities in units of $10^{12}\, \rm{cm}^{-2}$.  The average abundances for each velocity range appear in Table 5, computed relative to molecular hydrogen (columns 4 -- 7) and relative to hydrogen nuclei (columns 8 -- 11).  For this purpose, we adopted atomic hydrogen column densities, $N({\rm H)}$, and hydrogen nucleus column densities, $N_{\rm H} = N({\rm H}) + 2\,N({\rm H}_2)$, obtained with the assumptions given in the table footnote.  {The values given in Table 4 for the W51 $\velo_{\rm LSR}=62 - 75$~km/s window must be regarded with caution.  Material absorbing in this velocity interval may lie in close proximity to W51, where radiative excitation is important: except for the case of SH, the derived column densities, computed for the conditions in foreground diffuse clouds, may be unreliable.  For this reason, the W51 $\velo_{\rm LSR}=62 - 75$~km/s has not been included in Table 5.}

Our observations did not lead to any detection of H$_3$S$^+$.  The line emission spectrum is quite rich in this spectral region, and the presence of interloper emission lines (i.e.\ lines from species other than H$_3$S$^+$) reduces our sensitivity to broad H$_3$S$^+$ features.  {\bff In particular, an emission feature lying $\sim 7\,\rm km\,s^{-1}$ blueward of the H$_3$S$^+$ line -- most likely the $3_{21}-4_{14}$ transition of methanol at 293.464203~GHz -- greatly limits our sensitivity to $\rm H_3S^+$ emission from the background sources.}  However, good upper limits can be placed on the H$_3$S$^+$ optical depths associated with certain narrow $\rm H_2S$ absorption components.  The best limits on the optical depth ratio, $\tau({\rm H_3S^+})/\tau({\rm H_2S})$, are obtained for the 40~km/s absorption component in the foreground of W49N and the 30~km/s absorption component in the foreground of W31C.  The 3~$\sigma$ upper limits on $\tau({\rm H_3S^+})/\tau({\rm H_2S})$ obtained for these components are $1.06 \times 10^{-2}$ and $1.02 \times 10^{-2}$ respectively.  Figure A.6 in Appendix A shows the $\rm H_3S^+$ (green) and $\rm H_2S$ (red, scaled by a factor 0.1) optical depth spectra obtained for these two components, together with the 3~$\sigma$ upper limit on $\rm H_3S^+$ (blue).  These limits correspond to a column density ratio $N({\rm H_3S^+})/N({\rm H_2S}) \sim 0.028 (T_{\rm rot}/100\,\rm K).$  Here, $T_{\rm rot}$ is the rotational temperature for the metastable ($J=K$) states of (the symmetric top molecule) $\rm H_3S^+$, a negligible population being assumed for all other states within each $K$-ladder.

\section{Discussion}

\subsection{Measured column densities, abundances, and correlations among the molecules observed}

All the molecular abundances in Table 5 lie far below the solar abundance of sulphur, $(n_{\rm S}/n_{\rm H})_\odot = 1.3 \times 10^{-5}$ (Asplund et al.\ 2009), indicating that the four molecules together account for at most 0.14$\%$ of the sulphur nuclei within any of the velocity ranges we defined.  Moreover, {sulphur is notable in showing no measurable depletion in diffuse atomic clouds (Sofia et al.\ 1994), and only moderate levels of sulphur depletion (by a factor $\simlt 4$) have been inferred even in the dense ($n_{\rm H} \sim 10^5$~cm$^{-3}$) gas within the Horsehead PDR (Goicoechea et al. 2006)}, suggesting that the molecules that we have observed contain a similarly small fraction of the {\it gas-phase} sulphur nuclei.  

Although the chemical pathways leading to the formation of SH, H$_2$S, CS and SO are clearly inefficient in converting sulphur atoms and ions into sulphur-bearing molecules, they evidently operate in a rather uniform manner, with the relative abundances among the sulphur-bearing molecules showing a remarkable constancy.  In Figure 6, we show (top panels) the column densities of SH, CS and SO for each velocity interval in Table 3, as a function of $\rm N({H_2S})$.  The dependence is nearly linear in every case, with the column density ratios (bottom panels in Figure 6 and columns 6 -- 9 in Table 4) spanning a fairly narrow range; the $N({\rm SH})/N({\rm H_2S})$, $N({\rm CS})/N({\rm H_2S})$, and $N({\rm SO})/N({\rm H_2S})$ ratios lie in the ranges 1.12 -- 2.96, 0.32 -- 0.61, and 0.08 -- 0.30 respectively.  Furthermore, these column density ratios exhibit no obvious trend with $N({\rm H_2S})$ as the latter varies over a factor $\sim 13$.
In Figure 7, we present the column densities (left panel) and abundances (right panel) of SH, H$_2$S, CS and SO relative to H$_2$, as a function of $N({\rm H}_2)$.  Similar results are presented in Figure 8, but with abundances given relative to H nuclei as a function of $N_{\rm H}$.   Once again, there is no strong trend in abundance with either $N({\rm H}_2)$ or $N_{\rm H}$.

\subsection{Principal Component Analysis}
 
We have used Principal Component Analysis (PCA; e.g.\ Jolliffe 2002) to explore differences and similarities among the absorption spectra observed toward W31C and W49N.  These sources are suited particularly well to this analysis because their absorption spectra exhibit a complex structure involving multiple foreground velocity components, detected at high signal-to-noise ratio, and unassociated with the source itself.  In our analysis, we made use of the optical depths, as a function of LSR velocity, for the SH, H$_2$S, CS and SO transitions observed here, together with analogous results for seven additional transitions of interest: the HI 21 cm transition (Winkel et al.\ 2015); the CH $N=1, J=3/2-1/2$ lambda doublet (Gerin et al.\ (2010a); the H$_2$O$^+$ $1_{11}-0_{00}$ transition (Gerin et al. 2010b; Neufeld et al.\ 2010); the $^{13}$CH$^+$ $J=1-0$ transition (Godard et al. 2012); the SH$^+$ $J_N=1_2 - 0_1$ transition; the $\rm ArH^+$ $J=1-0$ transition (Schilke et al.\ 2014), and the CN $N=1-0$ transition (Godard et al.\ 2010).  In the case of those transitions (the SH, CH, SH$^+$, $^{13}$CH$^+$, and CN transitions) that show hyperfine structure, we used the hyperfine deconvolved spectra (i.e.\ the optical depths inferred for the strongest hyperfine component.)

These eleven ``optical depth spectra" were rebinned on a common velocity scale covering LSR velocities in the range 10 to 50~km/s for W31C and 35 to 75~km/s for W49N.  They are shown by the black histograms in Figure 9 (for W31C) and Figure C.1 in Appendix C for W49N.  Prior to implementing the PCA, we followed the standard procedure of scaling and shifting each optical depth spectrum so that the variances were unity and the means were zero.  This ensures that each spectrum is given equal weight.  The PCA method represents each optical depth spectrum as a linear combination of eleven orthogonal (i.e. uncorrelated) components, with the optical depth for the {\it i}th transition given by 
$$\tau_i(\velo) = a_i + b_i \sum_{j=1}^{11} C_{ij} S_j (\velo) ,\eqno(1)$$ 
where the $S_j (\velo)$ are the 11 principal component optical depth spectra shown in Figure 10 (for W31C) and Figure C.2 in Appendix C (for W49N), and $a_i$, $b_i$, and $C_{ij}$ are coefficients.  The $S_j (\velo)$ all have a mean of zero and a variance of unity, and the $C_{ij}$ obey the normalization condition $\sum_j C_{ij}^2 = 1$. 

A key feature of PCA is that the terms within the sum in equation (1) above make successively smaller contributions to the right-hand-side, with the first term accounting for most of the variance in the optical depth spectra, the second term the next largest fraction of the variance, and so on.  In this particular case, the first two principal components are found to account for 72\% of the total variance, and the red histograms in Figures 9 and B.1 show that a fairly good approximation to most of the observed spectra is obtained by retaining just the first two principal components\footnote{This feature of PCA, the ability to reduce the dimensionality of a dataset, is widely used in image compression.}.  The only notable exception is ArH$^+$, for which the first two terms in equation (1) yield a poor fit to the observed spectra; for that case, the {\it blue} histograms in Figures 9 and C.1 show the sum of the first {\it three} terms.

In Figure 11, we present, for each optical depth spectrum, $i$,  the first three coefficients, $C_{i1}$, $C_{i2}$, and $C_{i3}$ in the sum in equation (1).  The upper panel shows $C_{i2}$ versus $C_{i1}$, while the lower panel shows $C_{i3}$ versus $C_{i2}$.  Each transition occupies a specific position within these plots, which provides a valuable graphical representation of the similarities and differences between the spectra.  Color-coded squares show the coefficients for each transition, and lines are drawn from the origin to each point.  {The results plotted in Figure 11 were obtained from a simultaneous analysis of the W31C and W49N data, but similar results were obtained by analysing each source individually.}  Moreover, the relative positions of the SH, H$_2$S, CS, SO, CH, CN and HI transitions in the upper panel do not depend strongly upon whether the molecular ion transitions are included in the analysis.   Several features are noteworthy:

\begin{figure*}
\centering
\includegraphics[width=16 cm]{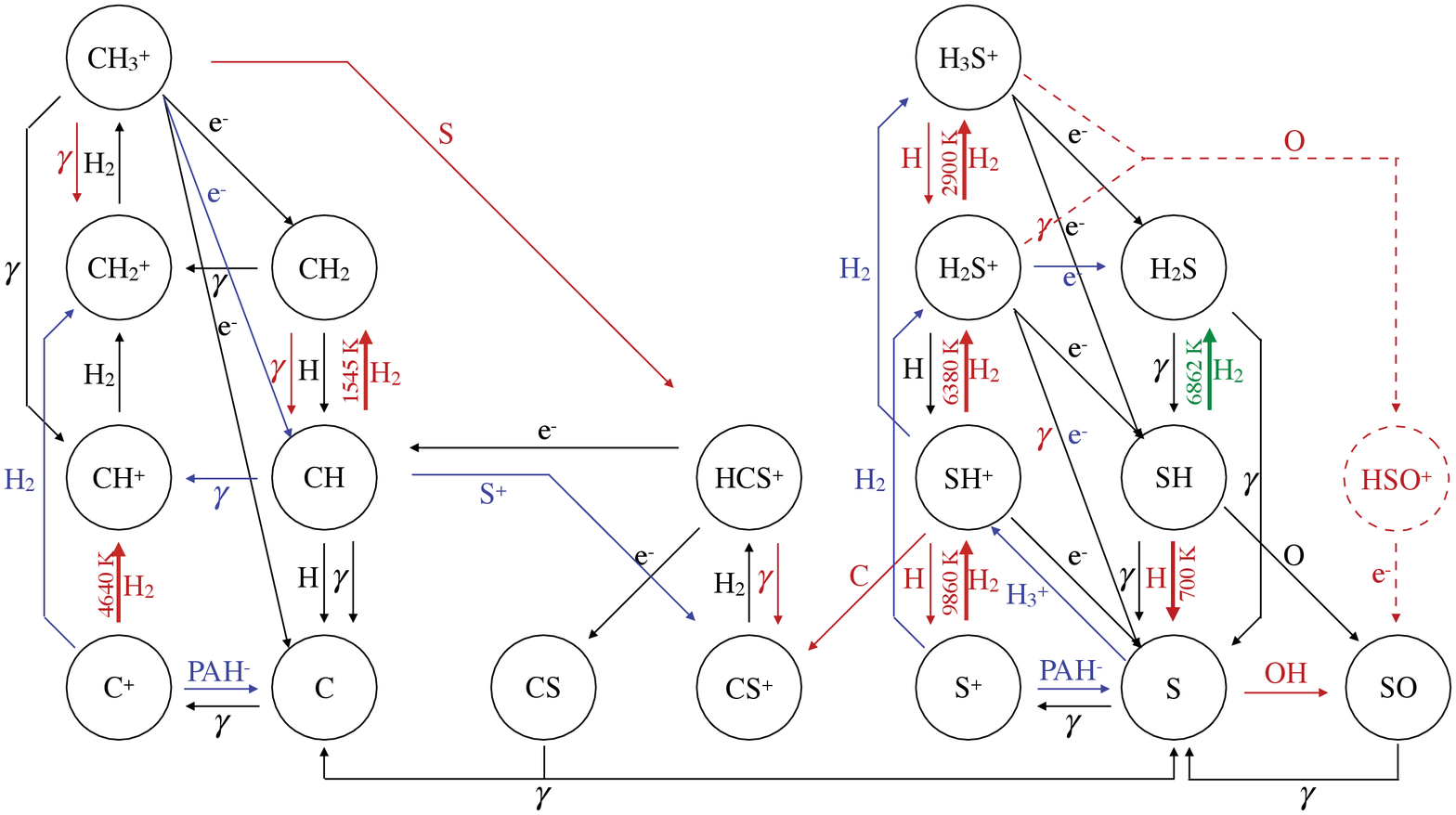}
\caption{Chemical network for sulphur- and carbon-bearing molecules (see text for details)} 
\end{figure*}

First, with the notable exception of ArH$^+$ transition, most transitions lie fairly close to the unit circle (black circle) in the upper panel, indicating that the first two principal components contain most of the information about their spectra.  (If the coefficients for the third and higher principal components were zero, then all points would lie exactly on the unit circle.)  ArH$^+$ is unusual in that its spectrum is largely accounted for by the {\it third} principal component.  The fact that it is poorly correlated with most of the other molecules may reflect its unusual chemistry (Schilke et al. 2014): it is believed to be a tracer of gas with an extremely small ($\simlt 10^{-3}$) molecular fraction.

Second, the transitions of the four {neutral} sulphur-bearing molecules lie very close to each other in the plot, are significantly separated from the HI, CH, and all the molecular ion transitions, and are relatively close to the CN transition.  In the limit where the points lie very close to the unit circle, the correlation coefficient between any pair of transitions is equal to cosine of the angle between the relevant vectors (i.e. the lines drawn from the origin to each point).  Indeed, the linear correlation coefficients among the neutral sulphur-bearing molecules range from 0.75 to 0.96 (with those among CS, H$_2$S, and SH lying in the range 0.90 to 0.96).  By contrast, the correlation coefficients {\it between} CH and the neutral sulphur-bearing molecules are smaller, in the range 0.52 and 0.67, and those between HI and the neutral sulphur-bearing molecules are smaller yet: 0.32 to 0.39.

\begin{figure}
\includegraphics[width=7.8 cm]{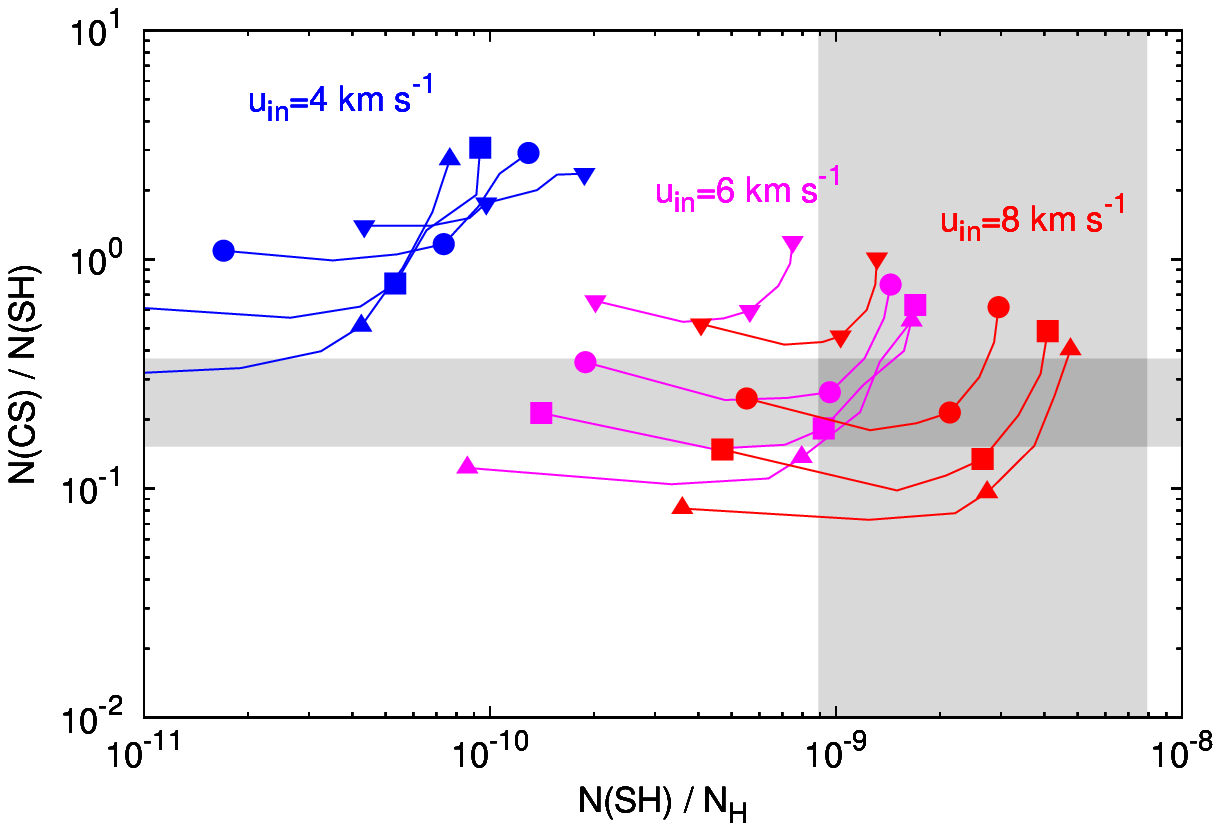}
\includegraphics[width=7.8 cm]{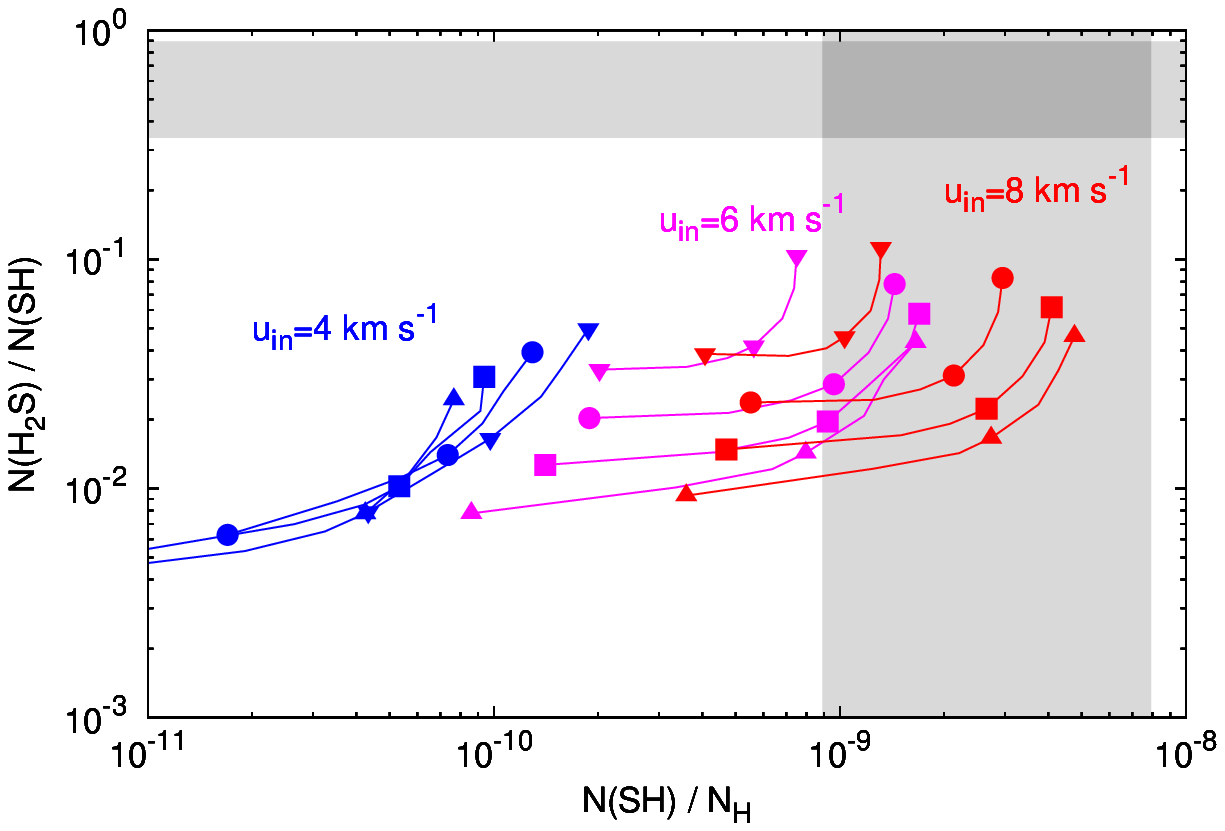}
\includegraphics[width=7.8 cm]{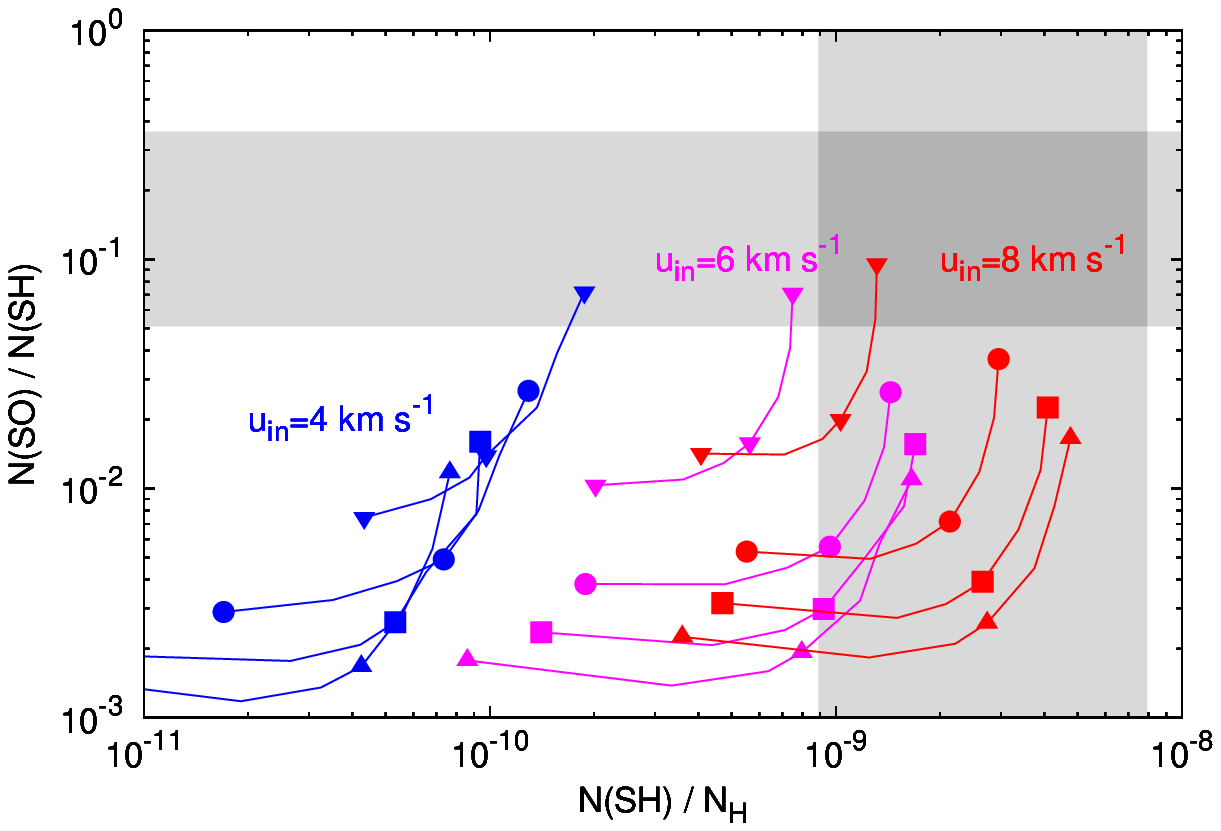}
\includegraphics[width=7.8 cm]{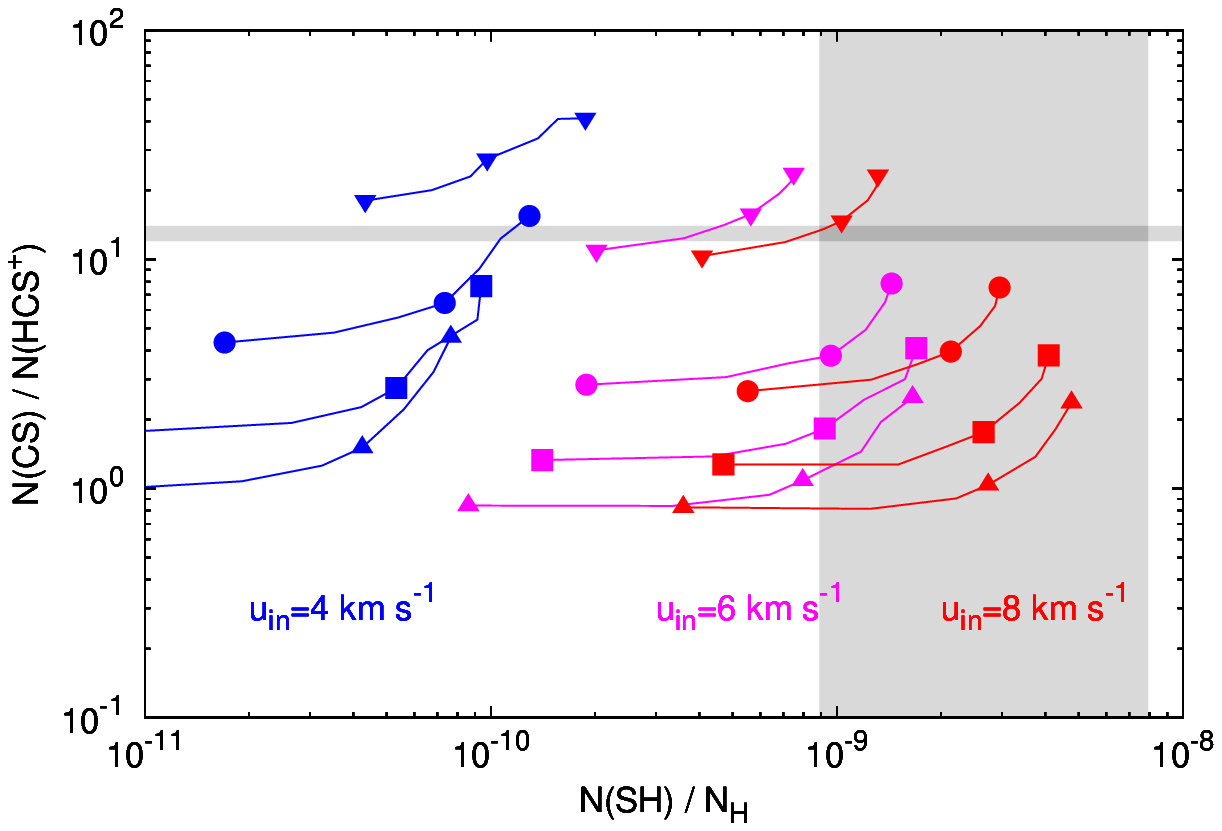}
\vskip 0.05 true in
\caption{Column density ratios predicted by standard TDR models, for a turbulent rate-of-strain of $3 \times 10^{-11} \,\rm s^{-1}$; ion-neutral velocity drifts of 4 (blue), 6 (magenta), and 8 (red) $\rm km \, s^{-1}$; and densities of 30 (triangles), 50 (squares), 100 (circles), and 300 (upside down triangles) cm$^{-3}$. The column densities are computed for a total hydrogen column density $N_{\rm H}$ of $1.8 \times 10^{21}\,\rm cm^{-2}$, assuming an homogeneous shielding $A_{\rm V}$ over the entire line-of-sight. Along each curve, $A_{\rm V}$ varies between 0.1 and 1.0 from left to right.  Grey shaded regions indicate the observed range of values.}
\end{figure}

Third, the molecular ions -- other than ArH$^+$ -- have transitions that typically lie close together and close to HI in the upper panel.  {In particular, the SH$^+$ molecular ion shows a distribution closer to that of CH$^+$ and H$_2$O$^+$ than to that of the neutral sulphur-bearing molecules observed in the present study.}  The molecular ions do, however, occupy a range of positions in the lower panel (i.e.\ they show a range of coefficients for the third principal component.)

Fourth, the vector corresponding to CH in Figure 11 lies {\it between} the vectors for HI and those for CN and the neutral sulphur-bearing molecules. CH is believed to be a good tracer for H$_2$, both in gas that is mainly atomic and in gas that is mainly molecular, while CN is thought to be abundant only in regions of large molecular fraction, so this behaviour may indicate that the neutral sulphur-bearing species are preferentially found in regions of large molecular fraction and are less prevalent in partially-molecular regions where CH is present {\bff but CN is absent.}  We note, in this context, that the relative positions of CH$^+$, CH, and CN in the upper panel of Figure 11 are consistent with previous UV absorption line observations (Lambert et al.\ 1990; Pan et al.\ 2004, 2005), which showed that CH line profiles could be seen as the sum of a broad component similar to the CH$^+$ profile, and a narrow one, similar to that of the CN line.

{\bff 
 
\subsection{Comparison with previous observations of CS, H$_2$S and SO}

Sulphur-bearing molecules have previously been studied within diffuse molecular clouds, both through observations of line emission (e.g.\ Drdla et al.\ 1989; Heithausen et 
al.\ 1998; Turner 1996a, 1996b) and absorption (Tieftrunk et al.\ 1994; Lucas \& Liszt 2002; Destree et al.\ 2009).  Absorption-line observations can provide the most reliable estimates of molecular column densities, because -- as noted in Appendix B -- the latter often depend only weakly upon the assumed physical conditions in the absorbing material or upon the adopted rate coefficients for collisional excitation.  Accordingly, we focus here on previous {\it absorption}-line observations, which provide the most direct quantitative comparison with results obtained in the present study.

Millimeter-wavelength observations of absorption by S-bearing molecules in diffuse clouds have been reported by Lucas \& Liszt (2002), who detected CS, H$_2$, and/or SO toward six extragalactic radio sources.  The typical $N({\rm CS})/N({\rm H}_2)$ ratio inferred from these observations was $4 \times 10^{-9}$, a value lying close to the middle of the range of $N({\rm CS})/N({\rm H}_2)$ obtained in the present study (Table 5).  Absorption line observations of CS in diffuse clouds are also possible at ultraviolet wavelengths through observations of the C--X (0,0) band near $1401 \AA$, which has been detected along 13 sight-lines toward hot stars (Destree et al.\ 2009).  The observed absorption line strengths are broadly consistent with the CS abundances inferred by Lucas \& Liszt (2002) and in the present study, but a detailed comparison is limited by lack of knowledge of the relevant $f$-values.

Lucas \& Liszt (2002) reported $N({\rm CS})/N({\rm SO})$ ratios in the range $1.7 \pm 0.8$, which overlaps the range ($2 - 6$) implied by Table 4, and $N({\rm CS})/N({\rm H_2S})$ ratios in the range $6 \pm 1$.  The H$_2$S column densities derived by Lucas \& Liszt (2002), however, appear to be based on a column-density-to-optical-depth ratio (their Table 2) that is a factor of $\sim 12$ too small; applying this correction brings their $N({\rm CS})/N({\rm H_2S})$ ratio down to $0.48 \pm 0.08$, in excellent agreement with the range ($0.32 - 0.61$) reported here in Table 4.
In addition, Lucas \& Liszt also detected the 85.348~GHz $J=2-1$ transition of HCS$^+$ in absorption toward one of the six sources, 3C111, with an implied $N({\rm CS})/N({\rm HCS^+})$ ratio of $13.3 \pm 0.1$.  (The HCS$^+$ $J=2-1$ frequency was not covered in the present study.) 

}

\subsection{Comparison with astrochemical models}

\subsubsection{Models without ``warm chemistry"}

Standard models of photodissociation regions (PDRs) are known
to greatly underpredict the abundances of sulfur-bearing molecules
(e.g. Tieftrunk et al.\ 1994).   {For the densities ($n_{\rm H} \sim \rm few \times 100$~cm$^{-3}$) and extinctions ($A_{\rm V} \sim \rm few \times 0.1\,mag$) typical of the diffuse foreground clouds we observed (e.g. Gerin et al.\ 2014), we find the abundances to be underpredicted by a factor of $\sim 100$,
even in models that include gas-grain chemistry (Vasyunin \& Herbst 2011) and the photodesorption and chemical desorption\footnote{Recent experiments by Dulieu et al.\ (2014) have suggested that the latter process is important in the case of water produced on bare grain surfaces.} of molecules from grain surfaces.}

\subsubsection{Models for turbulent dissipation regions}

As discussed by Neufeld et
al. (2012), the discrepancy described in the previous subsection 
may provide evidence for a warm
chemistry driven by turbulent dissipation in magnetized shocks
(e.g. Pineau des For\^ets et al.\ 1986) or in vortices (TDR model,
Godard et al.\ 2009) where the gas temperature and/or ion-neutral
drift velocity are sufficient to activate highly endothermic
formation pathways. Such a scenario has indeed proven
successful (Godard et al.\ 2014) in reproducing the large abundances
of CH$^+$ and SH$^+$ observed in the diffuse interstellar
gas assuming that turbulence dissipates in regions with density
$n_{\rm H} \le 50\,\rm  cm^{-3}$, ion-neutral drift velocity $u_{\rm in} \sim 3$~km/s, 
and at an average rate $\bar{\epsilon} = 10^{-24}\rm \, erg\,cm^{-3}\,s^{-1}$, 
a value in agreement with measurements of the kinetic energy transfer rate at large scales
(Hennebelle \& Falgarone 2012).

In Figure 12, we show the main production and destruction routes for 19 sulfur- and carbon-bearing molecules predicted by the PDR, TDR and shock models for a gas with density $n_{\rm H}=50\,\rm{cm}^{-3}$ and shielding $A_{\rm V}=0.4$~mag. For each species, only the processes that together contribute to more than 70\% of the total destruction and formation rates are displayed. Solid blue arrows indicate reactions that are only important in PDRs while solid red arrows indicate reactions that are only important in TDRs or shocks. Green and dashed red arrows denote additional processes discussed in \S 4.3.4 below. Activation energy barriers are indicated by temperatures ($\Delta E/k$) marked next to the relevant reaction arrow. The chemical network we used here is an updated version of that used in the Meudon PDR code (Le Petit et al. 2006, http://ism.obspm.fr).  {\bff Table D.1 in Appendix D lists the rates adopted for all the processes shown in Figure 12, together with those for two additional reactions for which updated rate coefficients have been used.}
Here, many of the rates adopted for endothermic reactions were obtained from those for the exothermic reverse reactions using the principle of detailed balance.  This procedure may introduce inaccuracies, because it is based upon the assumption that the reactants are in local thermodynamic equilibrium with the kinetic temperature.  At the low densities typically of relevance in astrophysical applications, the degree of vibrational excitation -- in particular -- may be highly subthermal.  Thus, the reaction rates for endothermic reactions with vibrationally-cold reactants may be significantly overestimated if vibrational excitation is substantially more efficient than translational energy in overcoming the energy barrier.  Unfortunately, detailed scattering calculations with a quantum chemistry potential are needed to determine which reactions are most affected; these are not typically available in the literature.

TDR model predictions are shown in Figure 13, where the predicted column density ratios 
are plotted against {\bff $N({\rm SH})/N_{\rm H}$.} 
Grey shaded regions indicate the range of observed values listed in Tables 4 and 5 (or, in the case of $N({\rm CS})/N({\rm HCS}^+)$, reported by Lucas \& Liszt 2002).  On the one hand, {\bff the observed $N({\rm CS})/N({\rm SH})$ and $N({\rm SH})/N_{\rm H}$ ratios} can be accounted for by models with an ion-neutral drift velocity of at least 6~km~s$^{-1}$, somewhat larger than that invoked to explain the CH$^+$/SH$^+$ column density ratio.  
This result suggests that turbulent energy dissipates over a distribution of
structures with different strengths rather than a single type of
event.  On the other hand, however, the TDR model predictions
systematically underestimate the column densities of both H$_2$S
and SO by roughly an order of magnitude unless {\rm $A_{\rm V}$ is unrealistically large ($A_{\rm V} \ge 1$).}\footnote{This confirms the {\it sense} of the discrepancy discussed by Neufeld et al.\ (2012).  However, Neufeld et al.\ (2012) overstated the magnitude of the discrepancy, because of a computational error that led to the observed SH/H$_2$S ratio being understated by a factor 10.}  {\bff Moreover, the CS/HCS$^+$ ratios reported by Lucas \& Liszt (2002) for the sight-line toward 3C111 are larger than the model predictions except for the highest assumed density ($n_{\rm H} = 300\,\rm \, cm^{-3}$).}

\subsubsection{Shock models}

A similar discrepancy is found when the predictions of standard shock models are compared with the data.  Using the model described by Lesaffre et al. (2013), we have conducted a parameter study in which the column densities of sulphur-bearing molecules behind a shock were computed as a function of the shock velocity, $\velo_s$, the preshock magnetic field, $B_{\rm 0}$, the preshock density, $n_{\rm H}$, and the UV shielding.  The latter is parameterized by the visual extinction, $A_{\rm V}$, to the cloud surface, which we assumed to be irradiated by the typical Galactic UV radiation field (Draine 1978).  As long as $\velo_s$ exceeds a value sufficient to drive the endothermic production reactions of relevance (about 10~km/s for $n_{\rm H} = 100\,\rm cm^{-3}$ and $A_{\rm V}=0.1$), the SH/H$_2$S ratio is virtually independent of $\velo_s$ and $B_{\rm 0}$.  As in the TDR case, the predicted H$_2$S and SO column densities relative {\bff to those of SH}
are underestimated by a factor of about 10 for clouds with $n_{\rm H} = 100\,\rm cm^{-3}$ {\bff and small shielding $A_{\rm V} \le 1$. Similarly, the CS/HCS$^+$ ratio is also underestimated except for high density values. Unlike TDR models, however, shock models underestimate the CS/SH ratio by a factor of about 5 for $A_{\rm V} \le 1$.}

\subsubsection{Additional effects on the chemistry}

As a possible explanation of the small observed $N({\rm SH})/N({\rm H_2S})$ ratios relative to the predictions of standard shock models, we have investigated two effects.  First, we considered the dependence of the predicted SH/H$_2$S column density ratio upon the branching ratio for dissociative recombination of H$_3$S$^+$.  In the standard model, we adopted a branching ratio for H$_2$S production of 0.17, the value measured by Kami{\'n}ska et al.\ (2008) for the production of D$_2$S by dissociative recombination of D$_3$S$^+$.  If we assume instead that every recombination of $\rm H_3S^+$ leads to the production of H$_2$S, the SH/H$_2$S ratio decreases to a value of about 5, i.e.\ about twice the typically-observed value.  

Second, we have investigated a physical effect not previously included in standard models for the chemistry of sulphur-bearing molecules in interstellar shock waves.  After being produced by dissociative recombination of $\rm H_2S^+$ or $\rm H_3S^+$, SH molecules will initially carry an imprint of the velocity of the ionized parents from which they formed; this velocity drift, relative to other neutral species, might significantly enhance the production of H$_2$S via the endothermic neutral-neutral reaction $\rm SH(H_2,H)H_2S$ (for which $\Delta E/k = 6862$~K).  As shown by Flower \& Pineau des For\^ets (1998), who followed
the velocity of CH molecules within a C-type shock, computing
the drift between SH and H$_2$ only requires the rate coefficient for momentum transfer between SH and H$_2$, $k_{\rm mt}({\rm SH,H_2})$.  In principle,
this rate could be obtained by subtracting the inelastic
collisions rates (including both reactive and non reactive processes)
from the total SH + H$_2$ collision rate. However, as far
as we know, the collision frequency for this system has never
been estimated; moreover the rates of elastic and inelastic processes
depend strongly on the gas kinetic temperature. As a
result, we have explored the shock model predictions over
a wide range of momentum transfer rate coefficient: $10^{-12} \le k_{\rm mt}({\rm SH,H_2}) \le 5 \times 10^{-10}\,\rm cm^3\, s^{-1}.$  

In Figure 14, we show the drift velocity of SH relative to H$_2$, as a function of distance through the shock.  As expected, the SH velocity lies somewhere between the ion velocity (red) and the H$_2$ velocity (black), with the exact behaviour depending upon the value adopted for $k_{\rm mt}({\rm SH,H_2})$. 

\begin{figure}
\includegraphics[width=8.7 cm]{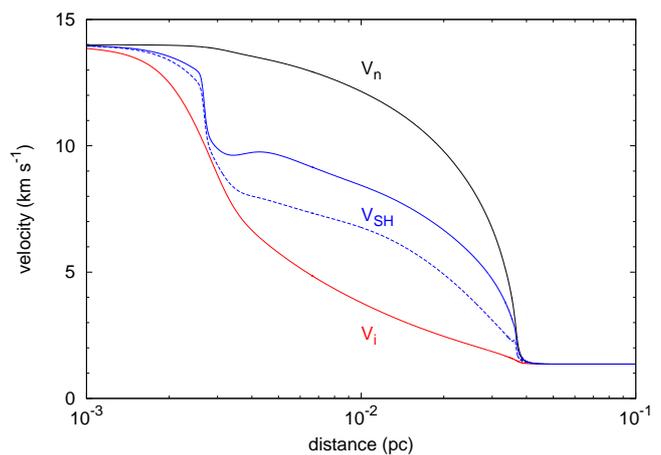}
\vskip 0.2 true in
\caption{Velocity of ions (red), SH (blue), and other neutral species (black), as a function of distance through the shock.   Results are presented for two values of the rate coefficient, $k_{\rm mt}({\rm SH,H_2})$, for momentum transfer between SH and H$_2$: $1 \times 10^{-10}$ (dashed blue curve) and  $5 \times 10^{-10}\,\rm cm^3\, s^{-1}$ (solid blue curve).} 
\end{figure}

\begin{figure}
\includegraphics[width=7.4 cm]{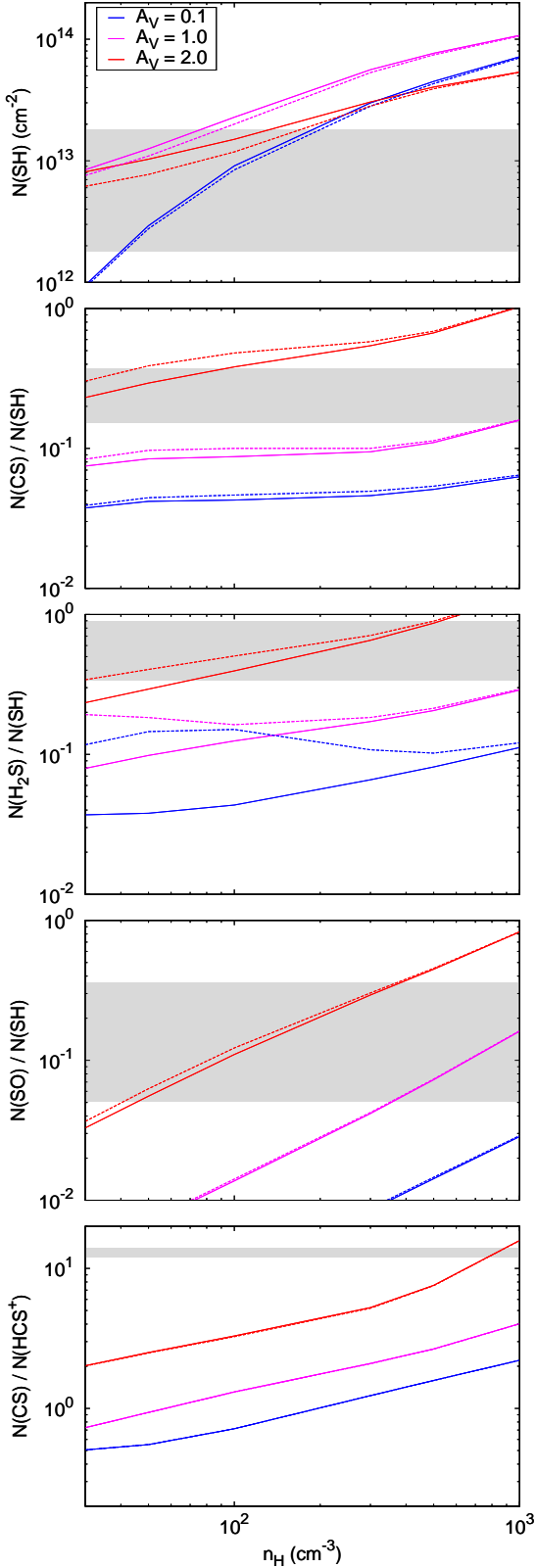}
\vskip 0.05 true in
\caption{Column density predictions for the standard shock model, as a function of preshock density, $n_{\rm H}$, and for $A_{\rm V}$ of 0.1 (blue), 1 (magenta), and 2 (red).  Velocity memory effects are included for SH for assumed values of $1 \times 10^{-10}\,\rm cm^3\, s^{-1}$ (dashed curves) and $5 \times 10^{-10}\,\rm cm^3\, s^{-1}$ (solid curves) for  $k_{\rm mt}({\rm SH,H_2})$.  Gray shaded regions indicate the observed range of values.}
\end{figure}
 
Figure 15 shows predictions for the shock model, with ``velocity memory effects" included for SH.  The reaction pathway enabled by such effects is shown by the green arrow in Figure 12.
The results are shown as a function of $n_{\rm H}$ and for three values of $A_{\rm V}$ (0.1, 1, and 2).  Here, the top panel shows the SH column density for a single shock, while the other {\bff four panels} show {\bff the CS/SH, H$_2$S/SH, SO/SH, and CS/HCS$^+$ column density ratios}, all computed for a single C-type shock with a velocity of 14~km/s and a preshock magnetic field $B_0=(n_{\rm H}/\rm cm^{-3})^{1/2}\, \mu$G.   Results are shown for $k_{\rm mt}({\rm SH,H_2})$ of $1 \times 10^{-10}\,\rm cm^3\, s^{-1}$ (dashed curves) and $5 \times 10^{-10}\,\rm cm^3\, s^{-1}$ (solid curves).  The grey shaded areas in each panel indicate the range of observed values presented in Table 4.  In the case of the top panel, the shaded area refers to the SH column density per magnitude of visual extinction; a comparison with the model predictions then leads to the number of shocks per magnitude of visual extinction required to reproduce the observations.  {\bff For $A_{\rm V} \le 1$}, the inclusion of velocity memory effects for SH, with an assumed $k_{\rm mt}({\rm SH,H_2})$ of $1 \times 10^{-10}\,\rm cm^3\, s^{-1}$ (dashed curves), significantly {\bff increases the predicted H$_2$S/SH ratio}, although the observed values still lie a factor $\sim 3$ {\bff above} the predictions unless $A_{\rm V}$ is unrealistically large ($\simgt 1$).  {\bff This process has however a small impact on all the other ratios displayed in Figure 15}.\footnote{\bff Predictions obtained for very small values of $k_{\rm mt}({\rm SH,H_2})$ smaller than  $3 \times 10^{-11}\,\rm cm^3\, s^{-1}$ (not shown) are in agreement with the observed H$_2$S/SH column density ratio even for very
diffuse ($n_{\rm H} \le 50\,\rm cm^{-2}$) and weakly shielded ($A_{\rm V}=0.1$) environments.} 
{\bff The SO/SH and CS/SH column density ratios are similarly underpredicted.}
The effect of velocity memory on the TDR model predictions will
be investigated in future studies but is expected to have an impact similar
to that found for shock models {\bff in increasing the predicted H$_2$S/SH column density ratio.}

Finally, as a possible explanation of the underprediction of
SO column densities by TDR and shock models, we have considered
a formation pathway of SO not yet included in standard
chemical networks and denoted by dashed red arrows in 
Figure 12: $\rm H_2S^+ + O \rightarrow HSO^+ + H$ followed by the
dissociative recombination of HSO$^+$. So far, standard chemical
networks available online (e.g. the UMIST database; McElroy et
al. 2013) only include the channels
$\rm H_2S^+ + O \rightarrow SH^+ + OH$ and $\rm H_2S^+ + O \rightarrow SO^+ + H_2$
with identical rate coefficients of $3.1 \times 10^{-10}\rm\, cm^3 \,s^{-1}$
(Prasad \& Huntress 1980), despite the fact that the production of HSO$^+$
is also a highly exothermic process. Implementing this channel
in the shock model with an assumed rate coefficient of
$6.2 \times 10^{-10}\rm\, cm^3 \,s^{-1}$ enhances
the predicted column density of SO by a factor of 10. Such a
scenario has yet to be confirmed by theoretical or experimental
estimates of the reactivity of $\rm H_2S^+$ with oxygen atoms and a
detailed analysis of the possible products.

\subsubsection{Line widths and velocity shifts}

{In the case of TDRs, an interpretation of the observed line profiles requires a convolution of the kinematic signatures of large scale turbulence with those of active regions of dissipation -- taking into account their orientations, strengths and distribution along the line of sight -- and of the relaxation stages that necessarily follow the dissipative bursts. In TDRs, SH$^+$, CH$^+$, and H$_3$S$^+$ are formed in the active phase (Godard et al. 2014), while SH, H$_2$S, SO, and CS are mainly produced by the relaxation phase. The linewidths of the former species are therefore expected to be wider than those of the latter; this effect is indeed apparent in the optical depth spectra (Figures 9 and C.1) and shows up in the PCA results (Figure 11).  In C-type shocks, the chemical production may induce a velocity shift between the line profiles of species. Flower \& Pineau des For\^ets (1998) estimate the shift between CH and CH$^+$ induced by a single shock of 14 km/s propagating along the line of sight to be roughly 2 km/s.  The same could be expected between SH and the neutral fluid, yet no significant line shifts are apparent in the observed data.  We note, however, that taking a random distribution of shocks with different orientations will drastically reduce the line shifts.  A detailed calculation of the line profiles in the framework of both the TDR and shock models remains to be done.}

{\section{Summary}

\noindent 1. With the use of the GREAT instrument on SOFIA, we observed the SH 1383~GHz $^2\Pi_{3/2} \, J = 5/2 - 3/2$ lambda doublet towards the star-forming regions W31C, G29.96--0.02, G34.3+0.1, W49N and W51.  We detected foreground absorption toward all five sources.
\vskip 0.1 true in

\noindent 2. Using the IRAM 30m telescope at Pico Veleta, we observed the H$_2$S $1_{10} - 1_{01}$ (169~GHz), CS $J=2-1$ (98~GHz), SO $3_2 - 2_1$ (99~GHz) and $\rm H_3S^+$ $1_0-0_0$ (293~GHz) transitions toward the same five sources. The $\rm H_3S^+$ $1_0-0_0$ transition was not detected in any source, but, for the other three transitions, we detected emission from the background source, and absorption by foreground clouds, toward every source.  
\vskip 0.1 true in

\noindent 3. In nine foreground absorption components detected towards these sources, the inferred column densities of the four detected molecules showed relatively constant ratios, with $N({\rm SH})/N({\rm H_2S})$ in the range 1.1 -- 3.0, $N({\rm CS})/N({\rm H_2S})$ in the range 0.32 -- 0.61, and $N({\rm SO})/N({\rm H_2S})$ in the range 0.08 -- 0.30.  The observed SH/H$_2$ ratios -- in the range 5 to 26 $\times 10^{-9}$ -- indicate that SH (and other sulphur-bearing molecules) account for $\ll 1 \%$ of the gas-phase sulphur nuclei. 
\vskip 0.1 true in

\noindent 4.  We have used Principal Component Analysis to investigate similarities and differences in the distributions of SH, H$_2$S, CS, SO, and seven other species present in foreground diffuse clouds along the sight-lines to W31C and W49N.   The neutral sulphur-bearing molecules we observed are very well correlated amongst themselves and with CN, moderately well correlated with CH (a surrogate tracer for H$_2$), and poorly correlated with atomic hydrogen and the molecular ions SH$^+$, CH$^+$, H$_2$O$^+$, and ArH$^+$.  A natural, but not unique, interpretation of these correlations is that SH, H$_2$S, CS, SO are preferentially found in material of large molecular fraction.   
\vskip 0.1 true in

\noindent 5.  The observed abundances of neutral sulphur-bearing molecules greatly exceed those predicted by standard models of cold diffuse molecular clouds, providing further evidence for the enhancement of endothermic reaction rates by elevated temperatures or ion-neutral drift. 
\vskip 0.1 true in

\noindent 6.  Standard models for C-type shocks or turbulent dissipation regions (TDRs) can successfully account for the observed abundances of SH and CS in diffuse clouds, but they fail to reproduce the column densities of H$_2$S and SO by a factor of about 10.  More elaborate shock models, which include ``velocity memory effects" for SH, reduce this discrepancy to a factor $\sim 3$.  Moreover, a possible reaction pathway not included in previous chemical networks, $\rm H_2S^+(O,H)HSO^+(e,H)SO$, can potentially explain the larger-than-expected SO abundances.
\vskip 0.1 true in

\noindent 7.  Further theoretical studies will be needed to determine the importance of velocity memory effects in TDRs and any kinematic signatures that might be expected.}

\begin{acknowledgements}
Based on observations made with the NASA/DLR Stratospheric Observatory for Infrared Astronomy. SOFIA Science Mission Operations are conducted jointly by the Universities Space Research Association, Inc., under NASA contract NAS2-97001, and the Deutsches SOFIA Institut under DLR contract 50 OK 0901.  This research was supported by USRA through a grant for Basic Science Program 01-0039. EF, MG and BG acknowledge support from the French CNRS/INSU Programme PCMI (Physique et Chimie du Milieu Interstellaire).  EH thanks NASA for its support of his research in the chemistry of the diffuse interstellar medium.  We thank B.~Winkel for communicating HI 21~cm absorption data prior to publication, and V.~Fish for providing data from Fish et al.\ (2003) in digital format.  {\bff We thank H.~Liszt for helpful discussions.}   We gratefully acknowledge the outstanding support provided by the SOFIA Operations Team, the GREAT Instrument Team, and the staff of the IRAM 30-m telescope.  

\end{acknowledgements}

\clearpage
\begin{appendix}
\section{Absorption features and molecular column densities} 

Figures A.1 through A.5 show the column density per velocity interval for each detected molecule in each observed source.  Figure A.6 presents upper limits on H$_3$S$^+$.

\begin{figure*}
\includegraphics[width=16 cm]{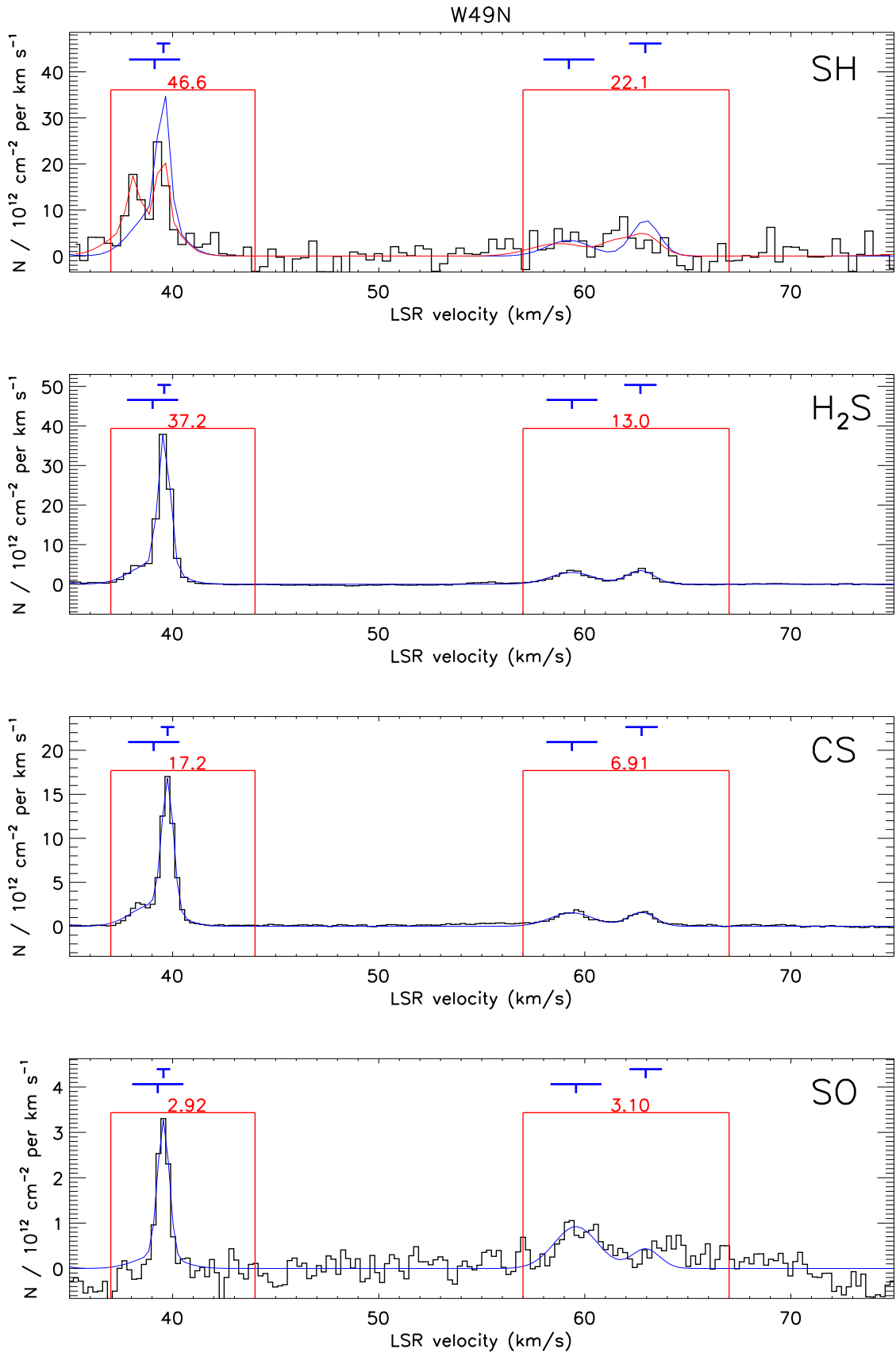}
\vskip 0.2 true in
\caption{Molecular column densities along the sight-line to W49N.  Black histograms show the data
and blue curves show a multi-Gaussian fit to the data.  In the case of the SH data, the red curve indicates a convolution of the multi-Gaussian fit with the hyperfine structure; it is this curve that should be compared with the black histogram (see text).  Red boxes indicate the velocity ranges
for which results are presented in Tables 1 and 2, with the red numbers at the top showing the integrated column density in units of $10^{12}\, \rm{cm}^{-2}$.  Blue symbols above each panel indicate the velocity centroid and FWHM for each Gaussian absorption component.}
\end{figure*}

\begin{figure*}
\includegraphics[width=16 cm]{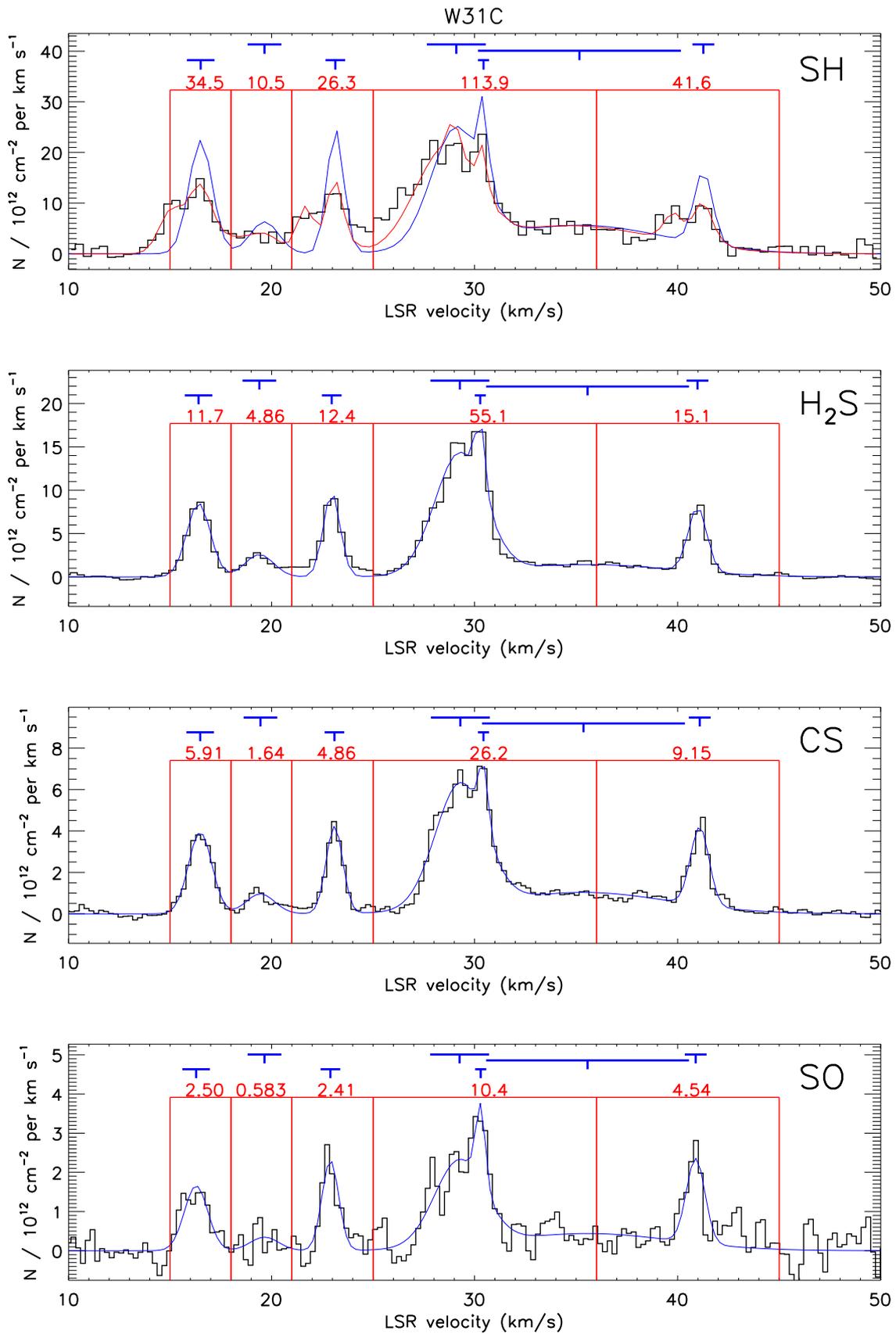}
\vskip 0.2 true in
\caption{Same as Figure A.1, but for W31C.}
\end{figure*}

\begin{figure*}
\includegraphics[width=16 cm]{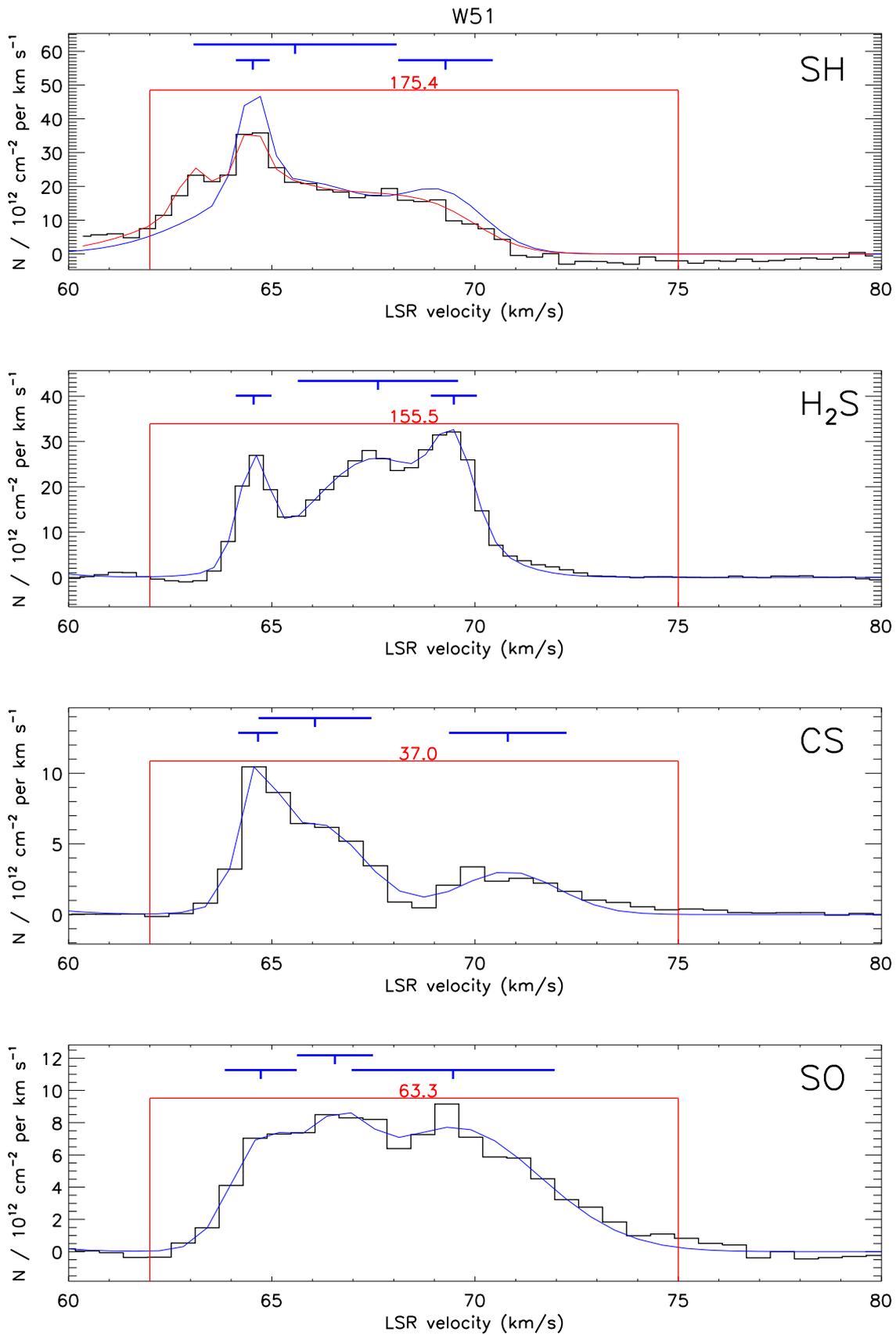}
\vskip 0.2 true in
\caption{Same as Figure A.1, but for W51.}
\end{figure*}

\begin{figure*}
\includegraphics[width=16 cm]{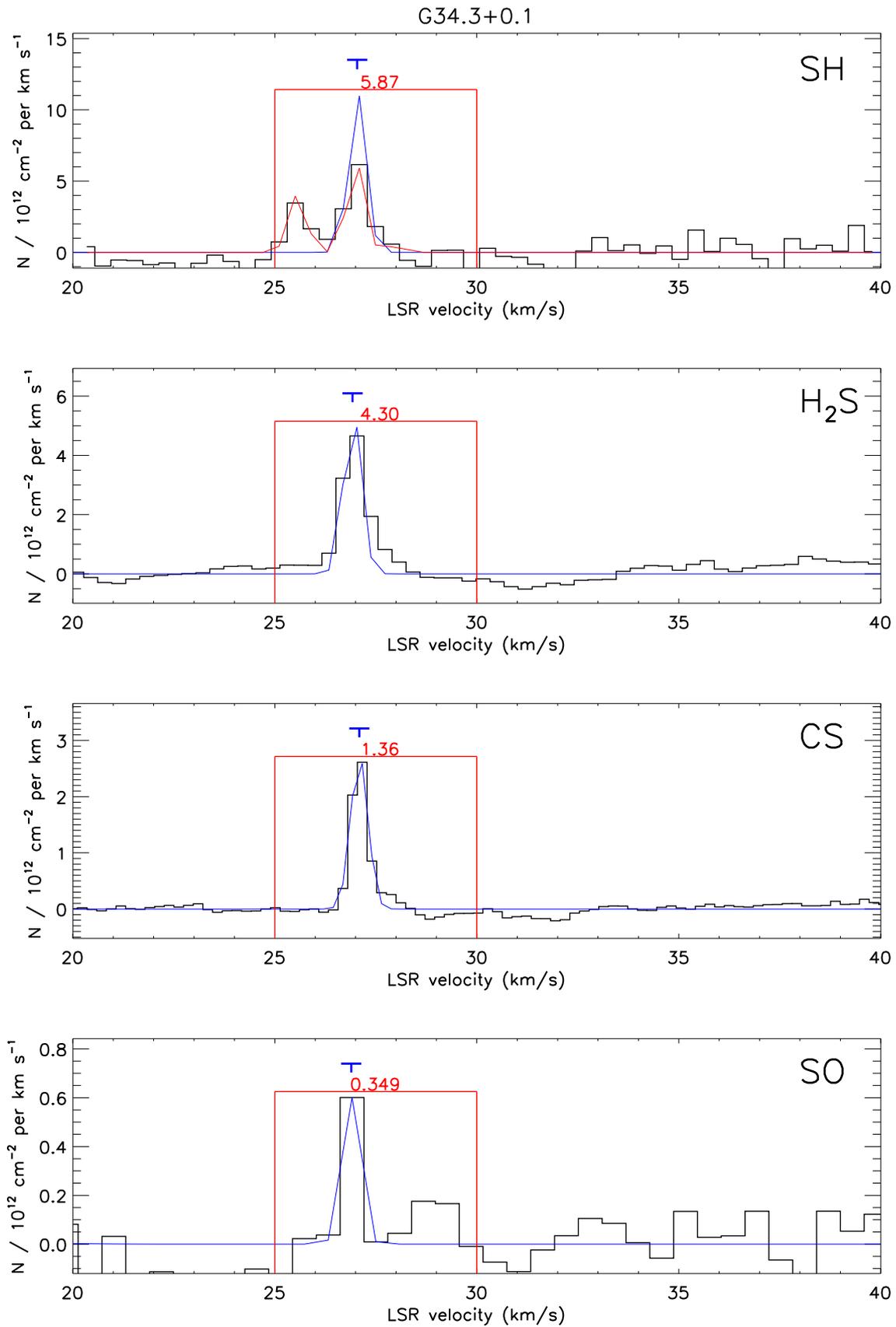}
\vskip 0.2 true in
\caption{Same as Figure A.1, but for G34.3+0.1.}
\end{figure*}

\begin{figure*}
\includegraphics[width= 16 cm]{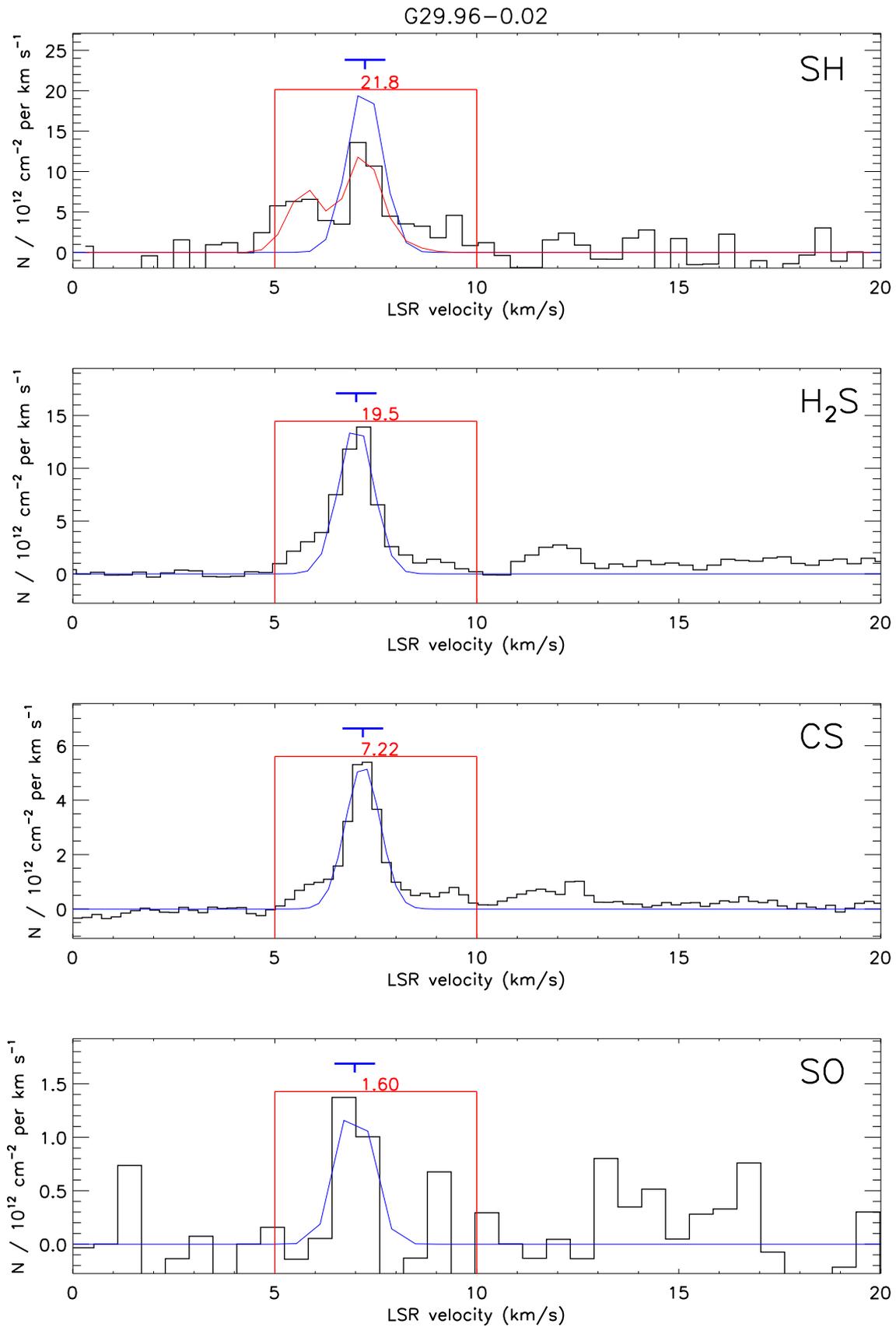}
\vskip 0.2 true in
\caption{Same as Figure A.1, but for G29.96-0.02.}
\end{figure*}

\begin{figure*}
\centering
\includegraphics[width= 13 cm]{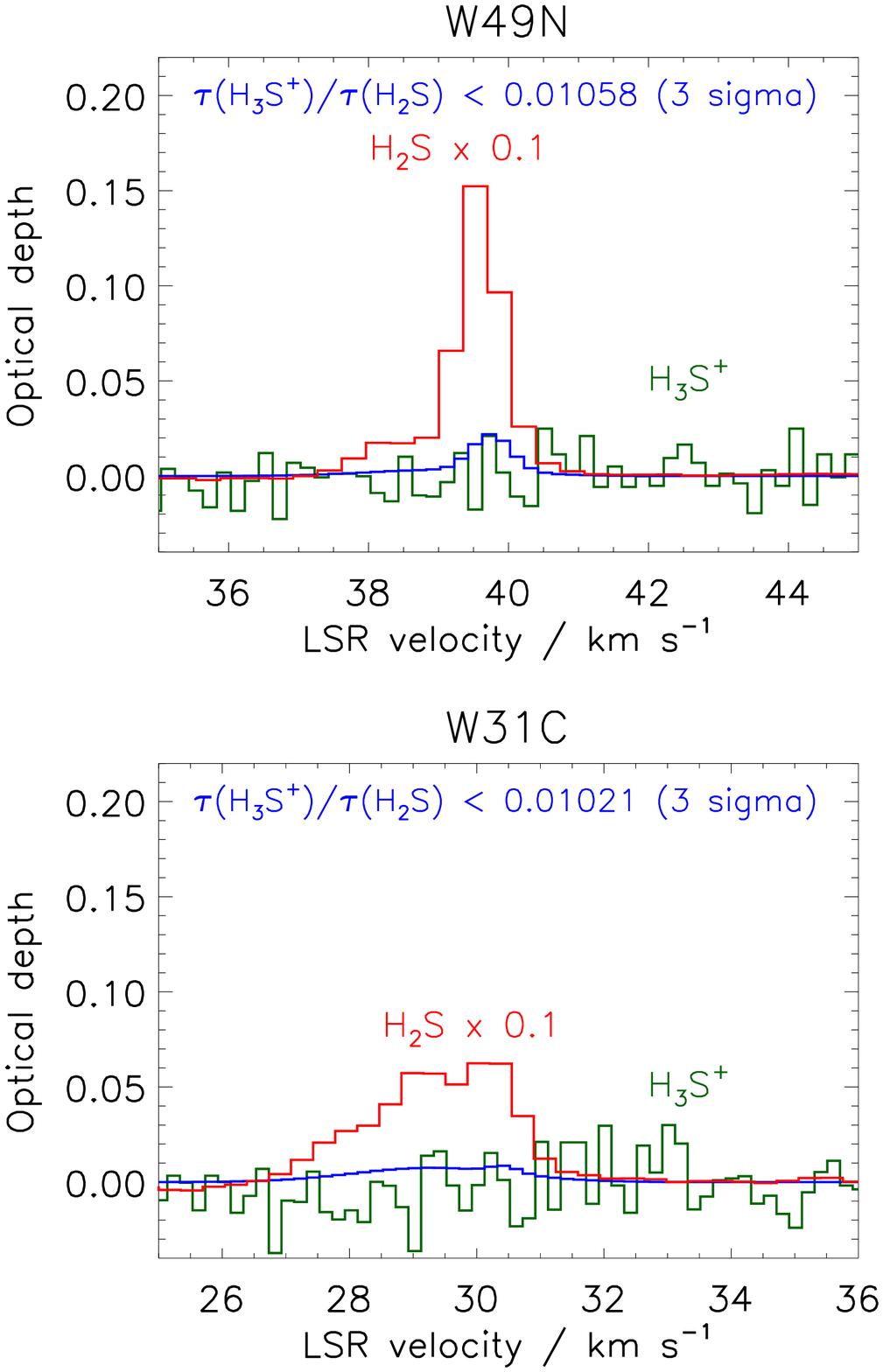}
\vskip 0.2 true in
\caption{$\rm H_3S^+$ (green) and $\rm H_2S$ (red, scaled by a factor 0.1) optical depth spectra obtained for the W49N 40~km/s foreground absorption component and the 30~km/s W31C foreground absorption component.  The $3\,\sigma$ upper limit on $\rm H_3S^+$ is shown in blue.}
\end{figure*}

\section{Excitation of sulphur-bearing molecules in diffuse molecular clouds}

In order to determine the equilibrium level populations for CS and SO in diffuse molecular clouds, we have solved the equations of statistical equilibrium as a function of temperature and density.  Our excitation model includes radiative pumping by the cosmic microwave background (CMB), at temperature $T_{\rm CMB} = 2.73$~K, spontaneous radiative decay, and collisional excitation by molecular hydrogen.  We adopted the collisional rate coefficients recommended in the LAMDA database (Sch\"oier et al.\ 2005)\footnote{http://$\rm home.strw.leidenuniv.nl/\sim moldata/$}, which were obtained by a simple scaling of the values computed for the excitation of CS (Lique et al.\ 2006a) and SO (Lique et al.\ 2006b) by He.  

In Figure B.1, we present results for the ratio of total molecular column density to optical depth, $N/\tau$, for the CS $J=2-1$ (upper panel) and SO $J_N = 3_2 - 2_1$ (lower panel) transitions that we observed.  Results are presented as a function of H$_2$ density for four different temperatures.  The left panels show the results on a log-log plot, while the right panels show the same data on a log-linear plot for a narrower range of $n({\rm H}_2)$.  
In the limit of low density, the level populations are in equilibrium with the CMB, at a temperature of 2.73~K, while in the limit of high density the level populations are in local thermodynamic equilibrium (LTE) at the gas temperature, $T$.  In the latter limit, $N/\tau$ is proportional to $T^2$ provided $kT \gg$ the transition energy, $\Delta E$ ; this result arises because the difference between the absorption and stimulated emission rates is proportional to  $$(N_l - N_u g_l/g_u) \propto N_l [1 - \exp(-\Delta E/kT)] \propto N_l/T  \propto N / (Z T) \propto T^{-2},$$ where $N_u$ and $N_l$ are the column densities in the lower and upper states, and $Z$ is the partition function, proportional to $T$ for diatomic molecules.   At intermediate densities and temperatures $\simgt 100$~K, the populations are inverted.

Two caveats affecting the results shown in B.1 are our neglect of radiative trapping and of  electron impact excitation in determining the level populations.   While the SO transition is clearly optically-thin in all the absorption components we observed, the optical depth of the CS transition can reach values of order unity at line center; thus, for CS, radiative trapping can lead to a modest reduction in the critical density at which departures from the low-density behaviour become significant.  For polar molecules like CS and SO, electron impact excitation can be of comparable importance to excitation by neutral species in regions where carbon is fully-ionized.  If the CS and SO absorption is occurring in regions where C is ionized, a further modest reduction in the critical density could result.

Notwithstanding these caveats, for the temperatures ($T \le 80~K$) and densities ($\le {\rm few} \times 10^{2} \rm \, cm^{-3}$)  typical of the foreground diffuse molecular clouds observed in this study (e.g. Gerin et al.\ 2014), the $N/\tau$ ratio is very comfortably in the low density limit for both CS and SO, justifying the assumption underlying our computation of column densities in \S3 above.  In the case of $\rm H_2$S, the effects of collisional excitation upon the $N/\tau$ ratio are expected to be even smaller than those for CS and SO, both because the 
spontaneous radiative rate for the $\rm H_2S$ transition is larger than those for CS and SO, and because the H$_2$S absorption takes place out of the {\bff lowest rotational state of the ortho spin-species}.

\begin{figure*}
\centering
\includegraphics[width= 16 cm]{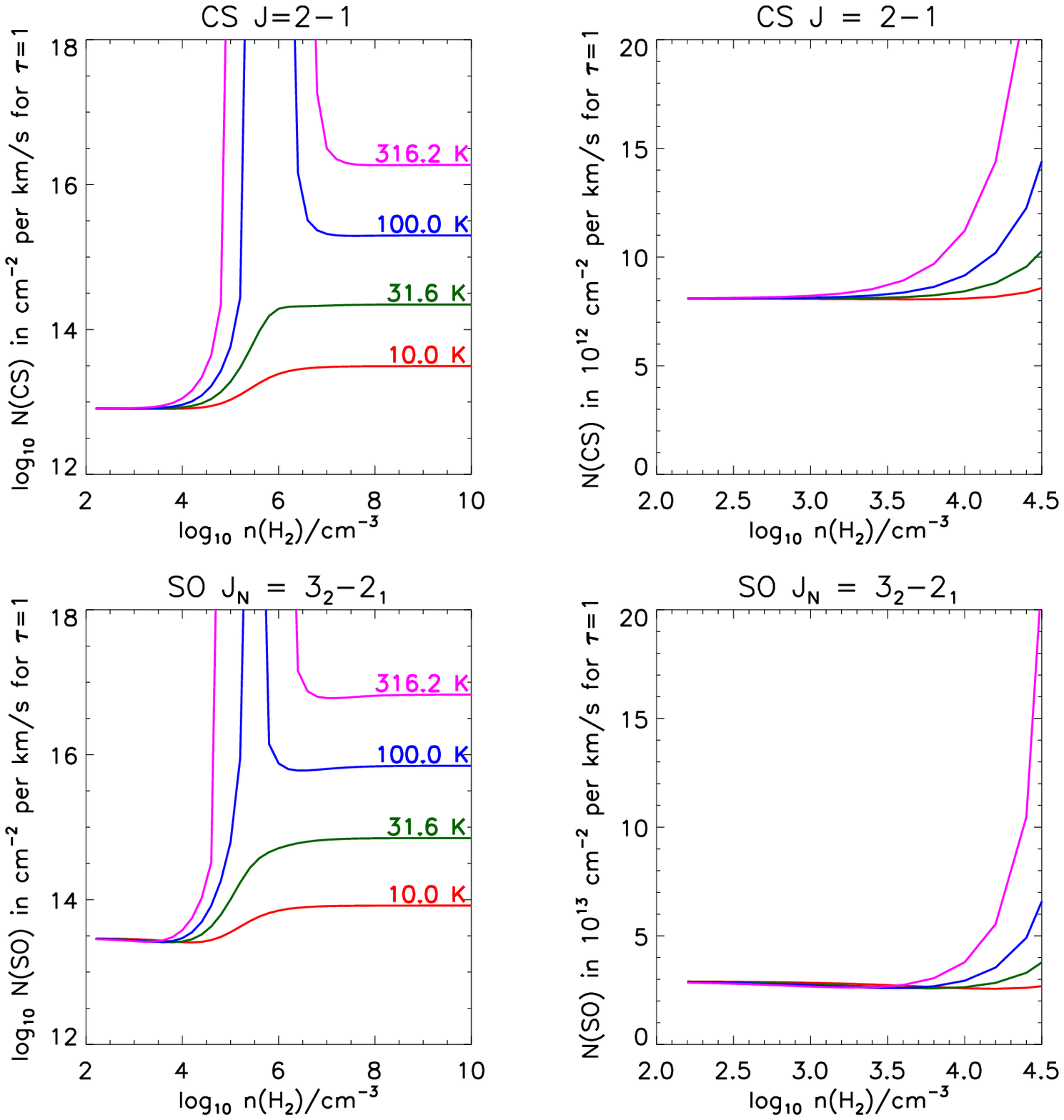}
\vskip 0.2 true in
\caption{Column densities required for an optical depth of unity.
}
\end{figure*}

\section{Principal component analysis for W49N}

Figures C.1 and C.2 show the optical depth spectra for eleven transitions observed towards W49N, and the principal components determined from a simultaneous analysis of the W31C and W49N spectra.

\begin{figure}
\includegraphics[width=9 cm]{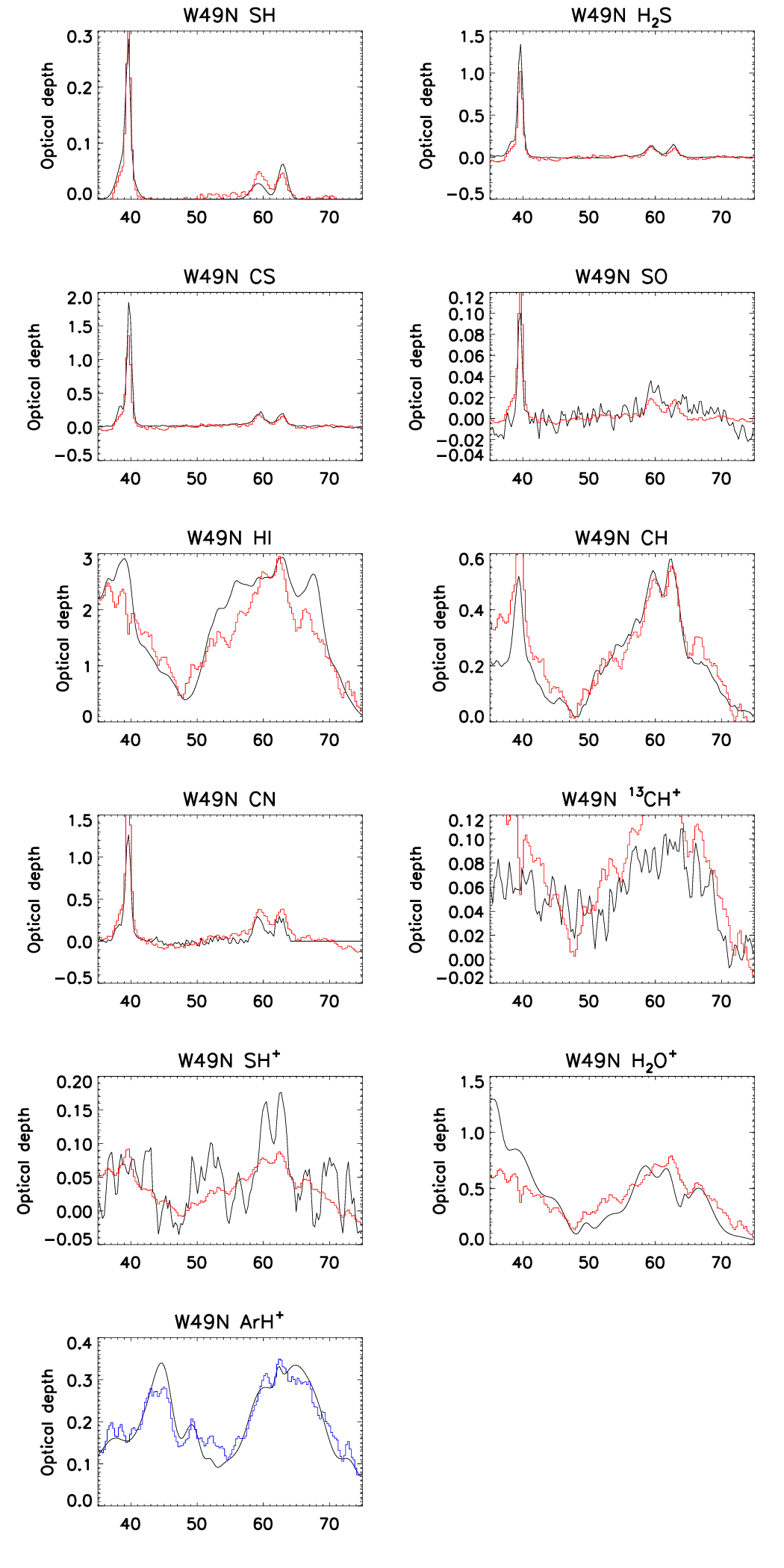}
\vskip 0.2 true in
\caption{Optical depths for eleven transitions observed toward W49N as a function of LSR velocity.  Black curve: rebinned observed data (see text).  Red histogram: approximate fit using only the first two principal components.  Blue histogram: approximate fit using only the first three principal components}
\end{figure}

\begin{figure}
\includegraphics[width=9 cm]{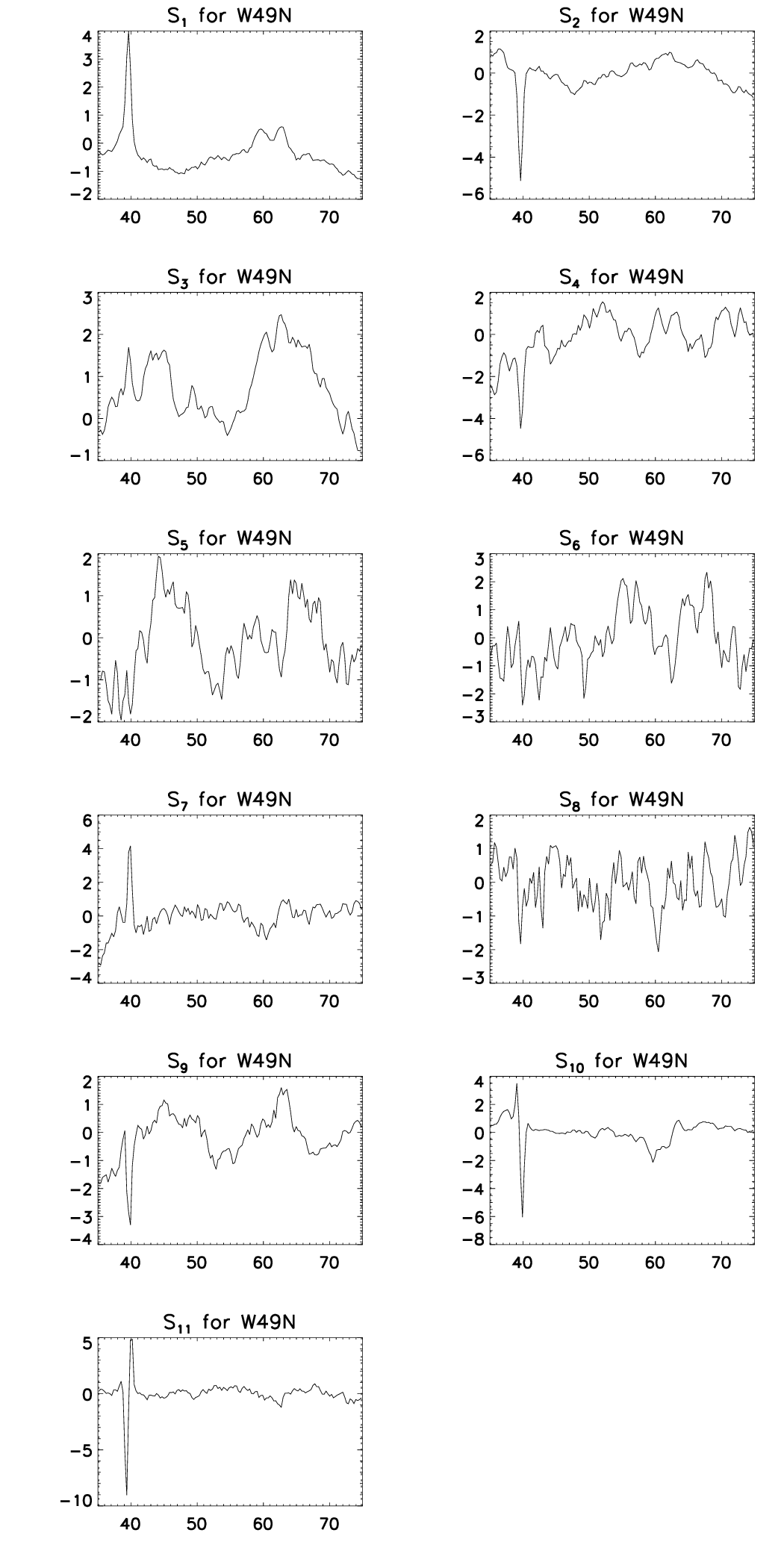}
\vskip 0.2 true in
\caption{Principal components obtained for W49N, from a joint analysis of the W31C and W49N absorption spectra}
\end{figure}

{\bff 

\section{Reaction network for sulphur-bearing molecules}

The chemical network for sulphur-containing molecules adopted in the Meudon PDR code version 1.4.4 -- a slightly-updated version of which is used here for TDR and shock modeling -- includes 432 reactions. In Table D.1, we present an abbreviated list containing rates for those reactions and protoprocesses that play a dominant role in the formation and destruction of sulfur bearing species (Figure 12) and/or whose rates have been updated for use in our study.
\clearpage

\begin{table*}
\caption{Abbreviated reaction list}
\begin{tabular}{llrrc}
\hline
\\
Reaction$^a$ & \phantom{000} $A$ & $B$ \phantom{0} & $C$ \phantom{000} & Reference$^c$ \\
\\
\hline
\\
\input{rtable1.tex}
\\
\hline
\\
Photoprocess$^b$ & \phantom{000}$\zeta_0$ & $\gamma$ \phantom{0} & & Reference$^c$\\
\\
\hline
\\
\input{rtable2.tex}
\\
\hline
\end{tabular}
\vskip 0.1 true in
\tablefoottext{a}{Reaction rate coefficient $k(T)= A (T/300{\rm K})^B \exp(-C\,{\rm K}/T)\,\rm cm^3 \,s^{-1}$}
\vskip 0.1 true in
\tablefoottext{b}{Rate $\zeta =  \zeta_0 \exp(-\gamma A_{\rm V}\rm /mag) \,s^{-1}$}
\vskip 0.1 true in
\tablefoottext{c}{References: (1) Manion et al.\ (2013); (2) Peng et al.\ (1999); (3) Demore et al.\ (1997); (4) Tsuchiya et al.\ (1994);
(5) Millar et al.\ (1986); (6)~Herbst et al.\ (1989); (7) Millar \& Herbst (1990); (8) Decker et al.\ (2001); (9) Prasad \& Huntress (1980);
(10) Hellberg et al.\ 2005; (11)~Abouelaziz et al.\ (1992); (12) Kaminska et al.\ (2008); (13) Montaigne et al.\ (2005); (14) Kim et al.\ (1974); 
(15) Anicich et al.\ (1975);
(16)~van~Dishoeck et al.\ (1988); (17) van Dishoeck (2006); (18) estimate}
\vskip 0.1 true in
\tablefoottext{d}{Value different from that adopted previously in the Meudon PDR code version 1.4.4}

\end{table*}

}
\end{appendix}
\end{document}

%% file: rms.tex
         W31C &       SH &   1383.2 GHz & 4.11 &  58.7 &  0.40 &    70 \\
     ...      &      H$_2$S  &            168.8 GHz & 0.75 &   3.8 &  0.35 &   198 \\
     ...      &           CS &             98.0 GHz & 0.81 &  10.8 &  0.24 &    75 \\
     ...      &           SO &             99.3 GHz & 0.78 &   9.4 &  0.24 &    83 \\
     ...      &   H$_3$S$^+$ &            293.5 GHz & 0.86 &  12.3 &  0.20 &    69 \\
\\
  G29.96-0.02 &       SH &   1383.2 GHz & 1.47 &  67.3 &  0.40 &    21 \\
     ...      &      H$_2$S  &            168.8 GHz & 0.36 &   3.9 &  0.35 &    92 \\
     ...      &           CS &             98.0 GHz & 0.42 &   6.4 &  0.24 &    65 \\
     ...      &           SO &             99.3 GHz & 0.43 &   5.5 &  0.59 &    78 \\
     ...      &   H$_3$S$^+$ &            293.5 GHz & 0.41 &  10.7 &  0.20 &    37 \\
\\
    G34.3+0.1 &       SH &   1383.2 GHz & 4.38 &  66.0 &  0.40 &    66 \\
     ...      &      H$_2$S  &            168.8 GHz & 1.47 &   4.6 &  0.35 &   320 \\
     ...      &           CS &             98.0 GHz & 1.32 &   5.9 &  0.24 &   222 \\
     ...      &           SO &             99.3 GHz & 1.35 &   5.0 &  0.59 &   269 \\
     ...      &   H$_3$S$^+$ &            293.5 GHz & 1.53 &  14.7 &  0.20 &   104 \\
\\
         W49N &       SH &   1383.2 GHz & 3.54 & 208.3 &  0.40 &    17 \\
     ...      &      H$_2$S  &            168.8 GHz & 1.46 &   5.3 &  0.35 &   278 \\
     ...      &           CS &             98.0 GHz & 1.57 &  11.2 &  0.24 &   140 \\
     ...      &           SO &             99.3 GHz & 1.54 &  12.7 &  0.24 &   120 \\
     ...      &   H$_3$S$^+$ &            293.5 GHz & 1.63 &  16.9 &  0.20 &    96 \\
\\
          W51 &       SH &   1383.2 GHz & 4.82 &  67.6 &  0.40 &    71 \\
     ...      &      H$_2$S  &            168.8 GHz & 0.96 &   7.9 &  0.35 &   120 \\
     ...      &           CS &             98.0 GHz & 0.62 &   6.9 &  0.60 &    90 \\
     ...      &           SO &             99.3 GHz & 0.62 &   6.9 &  0.59 &    90 \\
     ...      &   H$_3$S$^+$ &            293.5 GHz & 1.41 &  28.4 &  0.20 &    49 \\
\\

%% file: column.tex
        W49N & 37 -- 44     & $ \phantom{0} 46.6\, (7.7) $     & $ \phantom{0} 37.2\, (0.2) $     & $ \phantom{0} 17.2\, (0.1) $     & $\phantom{00}  2.9\, (0.3) $        &  $1.25\,\,\,\,\,\,(0.21)$                &  $0.46\,(< 0.01)$                &  $0.08\,(< 0.01)$ \\
        W49N & 57 -- 67     & $ \phantom{0} 22.1\, (8.3) $     & $ \phantom{0} 13.0\, (0.2) $     & $\phantom{00}  6.9\, (0.1) $     & $\phantom{00}  3.1\, (0.4) $        &  $1.70\,\,\,\,\,\,(0.63)$        &  $0.53\,\,\,\,\,\,(0.01)$        &  $0.24\,\,\,\,\,\,(0.03)$ \\
\\
        W31C & 15 -- 18     & $ \phantom{0} 34.5\, (1.6) $     & $ \phantom{0} 11.7\, (0.1) $     & $\phantom{00}  5.9\, (0.1) $     & $\phantom{00}  2.5\, (0.3) $        &  $2.96\,\,\,\,\,\,(0.14)$        &  $0.51\,\,\,\,\,\,(0.01)$        &  $0.21\,\,\,\,\,\,(0.02)$ \\
        W31C & 18 -- 21     & $ \phantom{0} 10.5\, (1.2) $     & $\phantom{00}  4.9\, (0.1) $     & $\phantom{00}  1.6\, (0.1) $     & $\phantom{00}  0.6\, (0.2) $        &  $2.17\,\,\,\,\,\,(0.26)$        &  $0.34\,\,\,\,\,\,(0.02)$        &  $0.12\,\,\,\,\,\,(0.05)$ \\
        W31C & 21 -- 25     & $ \phantom{0} 26.3\, (1.4) $     & $ \phantom{0} 12.4\, (0.2) $     & $\phantom{00}  4.9\, (0.1) $     & $\phantom{00}  2.4\, (0.3) $        &  $2.13\,\,\,\,\,\,(0.12)$        &  $0.39\,\,\,\,\,\,(0.01)$        &  $0.20\,\,\,\,\,\,(0.02)$ \\
        W31C & 25 -- 36     & $            113.9\, (2.4) $     & $ \phantom{0} 55.1\, (0.3) $     & $ \phantom{0} 26.2\, (0.2) $     & $ \phantom{0} 10.4\, (0.5) $        &  $2.07\,\,\,\,\,\,(0.05)$                &  $0.48\,(< 0.01)$                &  $0.19\,(< 0.01)$ \\
        W31C & 36 -- 45     & $ \phantom{0} 41.6\, (2.3) $     & $ \phantom{0} 15.1\, (0.2) $     & $\phantom{00}  9.1\, (0.2) $     & $\phantom{00}  4.5\, (0.4) $        &  $2.76\,\,\,\,\,\,(0.16)$        &  $0.61\,\,\,\,\,\,(0.02)$        &  $0.30\,\,\,\,\,\,(0.03)$ \\
\\
         W51 & 62 -- 75$^b$     & $            175.4\, (2.1) $     & $            155.5\, (1.7) $     & $ \phantom{0} 37.0\, (0.5) $     & $ \phantom{0} 63.3\, (1.1) $        &  $1.13\,\,\,\,\,\,(0.02)$                &  $0.24\,(< 0.01)$                &  $0.41\,(< 0.01)$ \\
\\
   G34.3+0.1 & 25 -- 30     & $\phantom{00}  5.9\, (1.1) $     & $\phantom{00}  4.3\, (0.3) $     & $\phantom{00}  1.4\, (0.1) $  & $         <  1.2\,(3\,\sigma) $        &  $1.36\,\,\,\,\,\,(0.28)$        &  $0.32\,\,\,\,\,\,(0.03)$             & $< 0.3\,(3\,\sigma)$ \\
\\
 G29.96-0.02 &  5 -- 10     & $ \phantom{0} 21.8\, (4.1) $     & $ \phantom{0} 19.5\, (0.9) $     & $\phantom{00}  7.2\, (0.2) $     & $\phantom{00}  1.6\, (0.5) $        &  $1.12\,\,\,\,\,\,(0.22)$        &  $0.37\,\,\,\,\,\,(0.02)$        &  $0.08\,\,\,\,\,\,(0.03)$ \\
\\

%% file: abundance.tex
        W49N & 37 -- 44 & 1.76 &  2.35                     & 26.5                   & 21.2                   &  9.8                   &  1.7                   & 7.94                   & 6.34                   & 2.92                   & 0.50 \\
        W49N & 57 -- 67 & 4.19 &  4.30                     &  5.3                   &  3.1                   &  1.7                   &  0.7                   & 1.74                   & 1.03                   & 0.55                   & 0.24 \\
\\
        W31C & 15 -- 18 & 2.34 &  1.23                     & 14.7                   &  5.0                   &  2.5                   &  1.1                   & 5.84                   & 1.97                   & 1.00                   & 0.42 \\
        W31C & 18 -- 21 & 1.57 &  1.18                     &  6.7                   &  3.1                   &  1.0                   &  0.4                   & 2.44                   & 1.12                   & 0.38                   & 0.13 \\
        W31C & 21 -- 25 & 2.49 &  2.13                     & 10.6                   &  5.0                   &  2.0                   &  1.0                   & 3.71                   & 1.74                   & 0.68                   & 0.34 \\
        W31C & 25 -- 36 & 8.19 & 10.35                     & 13.9                   &  6.7                   &  3.2                   &  1.3                   & 4.26                   & 2.06                   & 0.98                   & 0.39 \\
        W31C & 36 -- 45 & 4.94 & 37.12                     &  8.4                   &  3.0                   &  1.9                   &  0.9                   & 0.89                   & 0.32                   & 0.19                   & 0.10 \\
\\
   G34.3+0.1 & 25 -- 30 & 1.07 &  3.72                     &  5.5                   &  4.0                   &  1.3   & $< 1.1\,(3\,\sigma)$                   & 1.00                   & 0.73                   & 0.23   & $<0.21\,(3\,\sigma)$ \\
\\
 G29.96-0.02 &  5 -- 10 & 1.43 &  1.11                     & 15.3                   & 13.6                   &  5.1                   &  1.1                   & 5.50                   & 4.91                   & 1.82                   & 0.40 \\
\\

%% file: rtable1.tex
   $\rm        S      +      H_2    \rightarrow       SH      +        H    $ & $  4.20 \times 10^{-10}$  &   0.00  & 10494.00 &     (1)$^d$  \\
   $\rm       SH      +      H_2    \rightarrow     H_2S      +        H    $ & $  7.36 \times 10^{-13}$  &   2.31  &  6862.00 &         (2)  \\
   $\rm       SH      +        H    \rightarrow        S      +      H_2    $ & $  1.39 \times 10^{-10}$  &  -0.18  &   697.00 &         (1)  \\
   $\rm        S      +       OH    \rightarrow       SO      +        H    $ & $  6.59 \times 10^{-11}$  &   0.00  &     0.00 &         (3)  \\
   $\rm       SH      +        O    \rightarrow       SO      +        H    $ & $  1.58 \times 10^{-10}$  &   0.00  &     0.00 &       (3;4)  \\
   $\rm      S^+      +      H_2    \rightarrow     SH^+      +        H    $ & $  2.20 \times 10^{-10}$  &   0.00  &  9860.00 &     (5)$^d$  \\
   $\rm      S^+      +      H_2    \rightarrow   H_2S^+      +     h\nu    $ & $  1.00 \times 10^{-17}$  &  -0.20  &     0.00 &         (6)  \\
   $\rm     SH^+      +      H_2    \rightarrow   H_2S^+      +        H    $ & $  1.90 \times 10^{-10}$  &   0.00  &  6380.00 &     (5)$^d$  \\
   $\rm     SH^+      +      H_2    \rightarrow   H_3S^+      +     h\nu    $ & $  2.40 \times 10^{-16}$  &  -0.80  &     0.00 &         (7)  \\
   $\rm     SH^+      +        H    \rightarrow      S^+      +      H_2    $ & $  1.10 \times 10^{-10}$  &   0.00  &     0.00 &         (5)  \\
   $\rm   H_2S^+      +      H_2    \rightarrow   H_3S^+      +        H    $ & $  1.40 \times 10^{-11}$  &   0.00  &  2900.00 &     (5)$^d$  \\
   $\rm   H_2S^+      +        H    \rightarrow     SH^+      +      H_2    $ & $  2.00 \times 10^{-10}$  &   0.00  &     0.00 &         (5)  \\
   $\rm   H_3S^+      +        H    \rightarrow   H_2S^+      +      H_2    $ & $  6.00 \times 10^{-11}$  &   0.00  &     0.00 &         (5)  \\
   $\rm     CS^+      +      H_2    \rightarrow    HCS^+      +        H    $ & $  4.50 \times 10^{-10}$  &   0.00  &     0.00 &         (8)  \\
   $\rm      S^+      +       CH    \rightarrow     CS^+      +        H    $ & $  6.20 \times 10^{-10}$  &  -0.00  &     0.00 &     (9)$^d$  \\
   $\rm     SH^+      +        C    \rightarrow     CS^+      +        H    $ & $  9.90 \times 10^{-10}$  &   0.00  &     0.00 &         (9)  \\
   $\rm   CH_3^+      +        S    \rightarrow    HCS^+      +      H_2    $ & $  1.40 \times 10^{ -9}$  &   0.00  &     0.00 &         (9)  \\
   $\rm    H_3^+      +        S    \rightarrow     SH^+      +      H_2    $ & $  2.60 \times 10^{ -9}$  &   0.00  &     0.00 &         (9)  \\
   $\rm     SH^+      +        e    \rightarrow        S      +        H    $ & $  2.00 \times 10^{ -7}$  &  -0.50  &     0.00 &         (9)  \\
   $\rm   H_2S^+      +        e    \rightarrow       SH      +        H    $ & $  2.40 \times 10^{ -7}$  &  -0.72  &     0.00 &    (10)$^d$  \\
   $\rm   H_2S^+      +        e    \rightarrow        S      +      H+H    $ & $  2.40 \times 10^{ -7}$  &  -0.72  &     0.00 &    (10)$^d$  \\
   $\rm   H_2S^+      +        e    \rightarrow     H_2S      +     h\nu    $ & $  1.10 \times 10^{-10}$  &  -0.70  &     0.00 &         (9)  \\
   $\rm   H_3S^+      +        e    \rightarrow       SH      +      H+H    $ & $  3.02 \times 10^{ -7}$  &  -0.50  &     0.00 &        (11)  \\
   $\rm   H_3S^+      +        e    \rightarrow     H_2S      +        H    $ & $  8.84 \times 10^{ -8}$  &  -0.50  &     0.00 &     (11;12)  \\
   $\rm   H_3S^+      +        e    \rightarrow       SH      +      H_2    $ & $  7.80 \times 10^{ -8}$  &  -0.50  &     0.00 &     (11;12)  \\
   $\rm   H_3S^+      +        e    \rightarrow        S      +    H_2+H    $ & $  5.20 \times 10^{ -8}$  &  -0.50  &     0.00 &     (11;12)  \\
   $\rm    HCS^+      +        e    \rightarrow       CH      +        S    $ & $  7.86 \times 10^{ -7}$  &  -0.57  &     0.00 &        (13)  \\
   $\rm    HCS^+      +        e    \rightarrow       CS      +        H    $ & $  1.84 \times 10^{ -7}$  &  -0.57  &     0.00 &        (13)  \\
   $\rm        O      +     H_2S    \rightarrow       SH      +       OH    $ & $  7.90 \times 10^{-12}$  &   1.00  &  1737.00 &     (1)$^d$  \\
   $\rm    H_3^+      +     H_2S    \rightarrow   H_3S^+      +      H_2    $ & $  3.70 \times 10^{ -9}$  &   0.00  &     0.00 & (14;15)$^d$  \\

%% file: rtable2.tex
   $\rm        S      +     h\nu    \rightarrow      S^+      +        e    $ & $  4.40 \times 10^{-10}$  &     2.58 &&       (16)  \\
   $\rm       SH      +     h\nu    \rightarrow        S      +        H    $ & $  6.50 \times 10^{-10}$  &     1.42 &&       (16)  \\
   $\rm     H_2S      +     h\nu    \rightarrow       SH      +        H    $ & $  1.00 \times 10^{ -9}$  &     1.87 &&    (16;17)  \\
   $\rm     H_2S      +     h\nu    \rightarrow        S      +      H_2    $ & $  1.00 \times 10^{ -9}$  &     1.87 &&    (16;17)  \\
   $\rm       CS      +     h\nu    \rightarrow        C      +        S    $ & $  6.30 \times 10^{-10}$  &     2.03 &&       (16)  \\
   $\rm       SO      +     h\nu    \rightarrow        S      +        O    $ & $  2.40 \times 10^{ -9}$  &     1.95 &&       (16)  \\
   $\rm    HCS^+      +     h\nu    \rightarrow     CS^+      +        H    $ & $  7.00 \times 10^{-10}$  &     1.70 &&       (18)  \\